\RequirePackage[2018-12-01]{latexrelease}
\documentclass[A4paper]{gGAF2e}

\usepackage{float}
\usepackage{graphicx}	
\usepackage{amsmath}	
\usepackage{amssymb}	

\newcommand\dd{\mathrm{d}}

\newcommand\ee{\mathrm{e}}
\newcommand\ii{\mathrm{i}}

\newcommand\f{\frac}
\newcommand\p{\upartial}

\newcommand\sech{\mathrm{sech}}

\begin{document}

\small

\jvol{00} \jnum{00} \jyear{2025} 

\markboth{Joshua J. Brown and Gordon I. Ogilvie}{Gravitational instability in a 3D disc}

\title{Nonlinear analysis of gravitational instability in a 3D gaseous disc}

\author{JOSHUA J. BROWN${\dag}$$^{\ast}$\thanks{$^\ast$Corresponding author. Email: jb2228@cam.ac.uk
\vspace{6pt}} and GORDON I. OGILVIE${\dag}$\\\vspace{6pt}${\dag}$Department of Applied Mathematics and Theoretical Physics,
University of Cambridge, Centre for Mathematical Sciences, Wilberforce Road, Cambridge CB3 0WA, UK}

\maketitle

\begin{abstract}
Astrophysical discs which are sufficiently massive and cool are linearly unstable to the formation of axisymmetric structures. In practice, linearly stable discs of surface density slightly below the threshold needed for this instability often form spiral structures, and can subsequently fragment or exhibit a state of self-sustained turbulence, depending on how rapidly the disc cools. This has raised the question of how such turbulence is possible in the linearly stable regime. We suggest a nonlinear mechanism for this phenomenon. We find analytically weakly nonlinear axisymmetric subcritical solitary equilibria which exist in linearly stable 3D discs that are close to the instability threshold. The energy of these ‘soliton’ solutions is only slightly higher than that of a uniform disc, and the structures themselves are expected to be unstable to non-axisymmetric perturbations. In this way, these subcritical solitary equilibria highlight a nonlinear instability and provide a possible pathway to a turbulent state in linearly stable discs.
\end{abstract}


\section{Introduction}

Discs consisting of gas in orbital motion around a massive central body are found in numerous situations in astronomy.  A smooth disc that is sufficiently massive and cool is unstable to the formation of structures.  The simplest form of gravitational instability (GI) consists of an axisymmetric, radially dependent perturbation in the surface density that grows exponentially in time.  This happens when the perturbation releases sufficient gravitational potential energy to offset the increases in internal and orbital energy that occur when entropy and angular momentum are conserved.  Non-axisymmetric forms of GI are also possible, such as the transient growth (`swing amplification') of shearing spiral density waves, which can additionally release kinetic energy from the orbital shear flow by transporting angular momentum outwards.  [See \citet{2016ARA&A..54..271K} for a review of GI in gaseous discs.]

In its simplest form the criterion for instability is $Q<1$, where $Q=\kappa c/\pi G\Sigma$ is the stability parameter of Toomre (1964). Here $\kappa$ is the epicyclic frequency of horizontal perturbations to circular orbital motion in the disc, and is equal to the orbital angular velocity $\Omega$ in the case of a Keplerian disc; $c$ and $\Sigma$ are the sound speed and the surface density of the gas. Thus $Q$ measures the ratio of the product of the stabilizing effects of rotation and pressure to the destabilizing effect of gravity. The criterion $Q<1$ applies strictly to infinitesimal adiabatic or isothermal axisymmetric perturbations (using the appropriate sound speed in each case) in a 2D model that neglects the thickness and structure of the disc in the vertical direction perpendicular to the plane of the disc.

The nonlinear outcome of GI in gaseous discs is typically found to be non-axisymmetric and to depend on thermal physics \citep{2001ApJ...553..174G}.  Most attention has been paid to the diabatic processes of heating and cooling.  If the gas can radiate away the heat generated in shocks resulting from the nonlinear spiral density waves in less than about half an orbital period, then the disc is found to fragment into bound objects, which could be the progenitors of stars or planets depending on the circumstances.  Otherwise a sustained gravitational turbulence of spiral density waves is obtained.

Interestingly, the average value of the stability parameter $Q$ in the turbulent state is significantly greater than the critical value, meaning that the system exhibits self-sustaining activity in a linearly stable regime, and raising the question of how the motion is maintained in a perfectly circular disc with no source of fluctuations.  While shearing spiral density waves are known to undergo some transient growth in this regime \citep[e.g.][]{1992MNRAS.256..685N}, a mechanism is needed to counteract the inexorable orbital shear and regenerate leading spiral waves from trailing ones. [Such mechanisms have been explored for collisionless stellar systems \citep{2022ARA&A..60...73S} but do not appear to apply to the gaseous discs we are interested in.]

In an attempt to begin to explore the nonlinear dynamical phase space of GI and to try to understand the self-sustaining behaviour, \citet{2022ApJ...934L..19D} investigated some nonlinear aspects of GI for an ideal fluid in a 2D local disc model. First considering the steady, axisymmetric problem, in which the only nonlinearity\footnote{The radial velocity in this model necessarily vanishes, along with the nonlinearity in the advective derivative.} is the adiabatic relation $P\propto\Sigma^\Gamma$ between (2D) pressure~$P$ and surface density~$\Sigma$, they found that the axisymmetric GI is subcritical for $\Gamma<5/3$ (or $\Gamma>2$) and that equilibria of finite amplitude then exist some way into the linearly stable regime $Q>1$. These equilibria can be periodic in the radial direction, but when computed in a large domain they have a tendency to become solitary, i.e.\ radially confined and therefore independent of boundary conditions which may be artificial. \citet{2022ApJ...934L..19D} showed further that these equilibria are unstable to non-axisymmetric perturbations in the form of trailing spiral density waves, thereby signposting a route towards the type of structure observed in numerical simulations.

The structures emerging from the 2D model generally have horizontal length-scales that are not much longer that the vertical scale-height of the disc. Concerns therefore arise that the 2D model may be inaccurate or even misleading. In this paper we set out to determine under what conditions the GI is subcritical for an ideal fluid in a 3D disc. We will consider the adiabatic relation $p\propto\rho^\gamma$ between (3D) pressure~$p$ and density~$\rho$ in an isentropic fluid and determine for which values of $\gamma$ the axisymmetric GI is subcritical. We will give an extended treatment for an isothermal gas ($\gamma=1$) as the technical details of the calculation are easier to follow in this case.

Before treating the weakly nonlinear problem, we must recognize that the linear stability problem for a 3D disc has been only sparsely treated in the existing literature. \citet{1965MNRAS.130...97G} calculated the vertical structure of polytropic (adiabatically stratified) self-gravitating discs and determined the conditions for marginal linear stability in the special cases $\gamma=1$, $\gamma=2$ and $\gamma=\infty$ (incompressible fluid). However, their results are not directly applicable to the problem of interest to us because they considered a uniformly rotating (rather than Keplerian) disc in which the vertical gravity is due entirely to the disc itself, neglecting the important contribution from the central object. More recently, \citet{2010MNRAS.406.2050M} calculated the vertical structure of polytropic, Keplerian discs, including the vertical gravity of both the disc and the central object, and computed axisymmetric modes in such discs. They determined the critical value of their stability parameter $Q_\text{3D}=\Omega^2/4\pi G\rho_\text{m}$ (involving the midplane density $\rho_\text{m}$) and the corresponding radial wavenumber as a function of the polytropic index, taking $\gamma=1.4$ throughout (which corresponds to a warm, diatomic gas); their discs are generally not adiabatically stratified.




There are some interesting connections between the problem we are considering and the work on dynamo theory with which Professor Andrew Soward, the dedicatee of this special issue, is most closely associated, and also with the wider context of fluid dynamics and nonlinear dynamical systems. A fluid dynamo is a flow in which the magnetic field is sustained or amplified by the motion of an electrically conducting fluid opposing the dissipative effects of electrical resistivity. Nonlinear dynamos such as those resulting from magnetic buoyancy instability, Tayler instability or the magnetorotational instability, in the absence of an imposed magnetic flux, involve the search for self-sustaining solutions in which the motions driven by a magnetic instability are responsible for sustaining the same magnetic field against dissipation. There exists a useful analogy between such nonlinear dynamos and self-sustaining solutions in shear flows such as pipe flow, which are linearly stable but can admit coherent states of finite amplitude, which are understood to play an important role in the transition to turbulence \citep{2008AN....329..750R}. Our work on GI also involves the computation of nonlinear equilibria of finite amplitude and is part of an attempt to understand how gravitational turbulence is sustained in a linearly stable regime $(Q>1)$. At another level, our study of GI involves the solution of Laplace's equation for the gravitational potential in the exterior of a thin disc, which relates closely to the exterior problem for the magnetic field in Andrew Soward's work on dynamos in thin discs \citep{1992GApFD..64..163S,1992GApFD..64..201S}.

The structure of this paper is as follows. In section \ref{s:2D}, we outline the equivalent weakly nonlinear 2D problem and solution. In section \ref{s:equilibria}, we formalise our assumptions and derive the equations governing 3D nonlinear equilibria, which we solve in section \ref{s:iso} in the isothermal case. In section \ref{timedep}, we examine the linear dispersion relation and reintroduce a slow time dependence to the nonlinear solutions, whose energy we study in section \ref{s:energy}. We present our results in section \ref{s:res}, discuss our findings in section \ref{s:discussion} and in section \ref{s:conclusion} we draw our conclusions.

\section{2D case: an instructive example}\label{s:2D}

The 2D case offers much insight and guidance which will keep us grounded whilst undertaking the 3D analysis. It will further be a useful point of comparison when we come to interpret our results. For this reason, we discuss briefly the key aspects of the nonlinear 2D problem, but note that more detailed treatments may be found in \citet{1984pgs2.book.....F} and \citet{2022ApJ...934L..19D}.

Consider the axisymmetric fractional surface density perturbation $\sigma = \tilde{\sigma}\ee^{\ii (k x - \omega t)}$, with $x$ a local radial coordinate. If we take only linear terms in the equations of motion, we obtain the linear dispersion relation \citep{1964ApJ...139.1217T}:
\begin{equation}\label{2dDR}
    \omega^2 = \kappa^2 - 2 \pi G \Sigma |k| + c^2 k^2,
\end{equation}
where $\kappa$ is the epicyclic frequency, $\Sigma$ the surface density, and $c$ the sound speed. This gives a band of unstable modes when $Q \equiv \frac{\kappa c}{\pi G \Sigma} < 1$, centred on the most unstable wavenumber $k_c = \frac{\pi G \Sigma}{c^2}$. Note that when $Q = 1$, $k_c = \kappa/c$.

At the onset of this instability, the system undergoes a stationary pitchfork bifurcation, with a criticality which depends on the physics adopted. This may be thought to happen generically due to the weakly nonlinear terms becoming important when the linear stabilisation/destabilisation is weak near the instability onset. These nonlinear terms themselves may be either stabilising or destabilising; either way, when they balance the linear terms, they allow for finite-amplitude equilibria close to the onset of instability.

Close to the bifurcation point, as anticipated by Ginzburg--Landau theory, these equilibrium disturbances comprise the critical Fourier mode modulated by a slowly varying envelope $A$, i.e.
\begin{equation}\label{sigmaansatz}
    \sigma = \varepsilon A(X,T)\cos(k_c x)
\end{equation}
for $X = \varepsilon x$ and $T = \varepsilon t$. (Here, $\varepsilon^2$ measures the small deviation of $Q$ from the bifurcation point). In this way, we have a solution which may thought to be comprised of a small band of wavenumbers centred on $k_c$. The equation governing the modulation is the nonlinear Klein--Gordon equation, first derived in this context by \cite{1979AZh....56..279M},
\begin{equation}\label{2DKGE}
    \frac{1}{c^2}\frac{\upartial^2 A}{\upartial T^2} -\frac{\upartial^2 A}{\upartial X^2} = -\alpha^2 A + \beta^2 A^3,
\end{equation}
\begin{equation}
    \text{where}\quad \alpha^2 = - 2 \frac{\kappa^2}{c^2} (Q^{-1})_2, \quad \beta^2 = \frac{(2 - \Gamma)(5-3\Gamma)}{8}\frac{\kappa^2}{c^2},
\end{equation}
and we identify $\Gamma = 3 - 2/\gamma$ as the appropriate adiabatic index for a self-gravitating 2D disc\footnote{A strongly self-gravitating disc has central (3D) pressure $p \propto g \Sigma$, where the gravitational field strength $g \propto \Sigma$ by Gauss' law. Consequently, $p \propto \Sigma^2 \propto \rho^\gamma$. For surface density $\Sigma \propto \rho H$ and 2D pressure $P \propto p H$, it follows that $P \propto \Sigma^{3-2/\gamma}$.} \citep{1972AnRFM..4..219A}.

The linear terms in the above equation may be seen to arise from the Taylor expansion of the dispersion relation (\ref{2dDR}) about the bifurcation point, that is, the onset of instability. This may be seen by letting $Q^{-1} = 1 + \varepsilon^2 (Q^{-1})_2$, $k = \kappa/c + \varepsilon \, \delta k$, and thinking of $\delta k$ as Fourier-conjugate to $X$. The final nonlinear term arises from the next-order quartic modification to the energy of the state. The weakly nonlinear effects are destabilising for $\Gamma<5/3$ and $\Gamma > 2$, and stabilising for $5/3<\Gamma< 2$. We are particularly interested in destabilising effects, when the bifurcation is subcritical, corresponding to the regime $\Gamma<5/3 \iff \gamma < 3/2$. We argue later that the slow development of the nonlinear equilibria is more faithfully modelled as an isothermal process, placing us firmly in the regime $\gamma < 3/2$.

Equation (\ref{2DKGE}) has the beautiful exact travelling soliton solution, identified by \cite{1979AZh....56..279M}:
\begin{equation}\label{2Dsoliton}
    A(X-uT) = \sqrt{2}\frac{\alpha}{\beta}\sech\left(\frac{\alpha\left(X - uT\right)}{\sqrt{1 - u^2/c^2}}\right)
\end{equation}
for arbitrary $|u|<c$, when $(Q^{-1})_2< 0$ and $\beta^2 > 0$. When $\beta^2 < 0$, supersonic travelling solitons are permitted in linearly unstable discs with $(Q^{-1})_2 > 0$. The existence of subsonic (including stationary) solitons in linearly stable discs is transient however, as they are themselves unstable, typically forming spiral structures which may then result in the formation of bound fragments or turbulence \citep{2022ApJ...934L..19D}.

\section{Equilibria in the local approximation}
\label{s:equilibria}

We now turn our attention to the associated 3D problem, in which the non-trivial vertical structure of the disc is explicitly captured. We work in the local approximation for astrophysical discs, also known as the shearing sheet, shearing box, etc. \citep[e.g.][]{2017MNRAS.472.1432L}. This well-known model is constructed around a reference point that follows the circular orbit of a test particle around the centre of the gravitational potential, and employs local Cartesian coordinates $(x,y,z)$ in the radial, azimuthal and vertical directions. The frame of reference rotates about the $z$-axis with the angular velocity $\Omega$ of the reference orbit.

We are looking for the local equivalent of axisymmetric equilibria that depend only on the radial ($x$) and vertical ($z$) coordinates and have no motion in those directions. By allowing for a dependence on $x$, we can find solutions that are structured in the radial direction on a length-scale comparable to the vertical thickness of the disc, but much smaller than the distance from the central object. An equilibrium can be described by its density $\rho(x,z)$, pressure $p(x,z)$, local gravitational potential $\phi(x,z)$ and azimuthal velocity perturbation $v_y(x,z)$. The total azimuthal velocity in the rotating reference frame is $u_y=v_y-Sx$, where $S$ is the rate of orbital shear, equal to $3\Omega/2$ in a Keplerian disc.

We assume that the equilibrium is symmetric about the midplane $z=0$ and occupies the region $-Z(x)<z<Z(x)$, surrounded by a vacuum, where $Z(x)$ is the semi-thickness of the disc. For an isothermal disc, formally $Z(x) = \infty$, although the density will be exponentially small at large $z$. The surface density is
\begin{equation}
  \Sigma(x)=\int_{-Z(x)}^{Z(x)}\rho(x,z)\,\dd z.
\end{equation}

The total gravitational potential consists of a local contribution $\phi(x,z)$, which is generated by the local disc material according to Poisson's equation,
\begin{equation}
  \f{\p^2\phi}{\p x^2}+\f{\p^2\phi}{\p z^2}=4\pi G\rho,
\end{equation}
and a global contribution, which is generated by the central object and the distant parts of the disc (as well as any halo or other component). When the global contribution to the gravitational potential is combined with the centrifugal potential associated with the rotating frame and expanded to second order about the reference point, we obtain the tidal potential
\begin{equation}
  \Phi_\text{t}=\f{1}{2}\nu^2z^2-\Omega Sx^2,
\end{equation}
where $\nu$ is the vertical oscillation frequency of particle orbits. This form of the tidal potential ensures that the family of circular particle orbits in the midplane $z=0$ is described locally by the azimuthal velocity field $-Sx$, for which the Coriolis force balances the gradient of $\Phi_\text{t}$. The epicyclic frequency $\kappa$ of horizontal oscillations of such orbits is given by $\kappa^2=2\Omega(2\Omega-S)$. We assume that $\kappa^2$ and $\nu^2$ are positive so that particle orbits are stable.

The relevant equations of mechanical equilibrium for the disc are the $x$ and $z$ components of the equation of motion,
\begin{align}
  -2\Omega v_y&=-\f{1}{\rho}\f{\p p}{\p x}-\f{\p\phi}{\p x},\\
  0&=-\f{1}{\rho}\f{\p p}{\p z}-\f{\p\phi}{\p z}-\nu^2z.
\end{align}

The simplest type of equilibrium is a horizontally invariant solution that depends only on $z$ and not on $x$. This `uniform disc' is the representation, in the local model, of a smooth, thin disc that varies in the radial direction only on a length-scale that is comparable to the radius, i.e.\ much longer than the vertical thickness and therefore not represented within the local model. Since we are concerned with the gravitational instability of such a smooth disc, we are interested in equilibria that are dynamically accessible from such a uniform solution.

We therefore consider a horizontally invariant equilibrium as a reference state. Fluid elements in the reference state can be labelled by their positions $(x_0,z_0)$ in the $xz$ plane. The reference state has density $\rho_0(z_0)$, pressure $p_0(z_0)$ and local gravitational potential $\phi_0(z_0)$. It also has $v_y=0$, meaning that gas has the same orbital motion as a test particle in the midplane. The reference state has uniform vorticity $2\Omega-S$ and surface density $\Sigma_0$.

In going from the reference state to the axisymmetric equilibrium, fluid elements move from $(x_0,z_0)$ to $(x,z)$, involving a 2D transformation of the $xz$ plane. In order to preserve mass, the density of the fluid changes according to
\begin{equation}
  \rho=\f{\rho_0}{J},
\end{equation}
where
\begin{equation}
  J=\begin{vmatrix}
    \dfrac{\p x}{\p x_0}&\dfrac{\p x}{\p z_0}\\[8pt]
    \dfrac{\p z}{\p x_0}&\dfrac{\p z}{\p z_0}
  \end{vmatrix}
\end{equation}
is the Jacobian determinant of the transformation.

Since we consider an ideal fluid problem, there is a free choice to be made regarding the relationship between pressure and density of the reference state. We consider a family of polytropic equilibria in which the pressure and density are related by a power law,
\begin{equation}
  p=K\rho^{1+1/n},
\end{equation}
where $K$ and $n$ are positive constants. If we further identify the adiabatic exponent $\gamma$ of our ideal fluid with the polytropic exponent $1 + 1/n$, then the specific entropy of the reference state is uniform. As the polytropic index $n$ (not generally an integer) varies from $0$ to $\infty$, we can examine a range of models from incompressible to isothermal. Since we consider an ideal fluid model, the specific entropy should be preserved and remain uniform under the transformation.\footnote{The assumption of uniform entropy instead of stable vertical stratification excludes baroclinic and buoyancy effects from our model. Relaxing this assumption yields a more complex problem, which may nonetheless be tackled with an approach similar to that presented in this paper, noting that in this case Ertel's potential vorticity (as well as the entropy) are preserved under the transformation.}

By introducing the specific enthalpy
\begin{equation}
  w=(n+1)K\rho^{1/n},
\end{equation}
we simplify the equations of equilibrium to
\begin{align}
  -2\Omega v_y&=-\f{\p\psi}{\p x},\label{eomx}\\
  0&=-\f{\p\psi}{\p z},\label{eomz}
\end{align}
where
\begin{equation}
  \psi=w+\phi+\f{1}{2}\nu^2z^2.
\end{equation}
It follows that $\psi$ depends only on $x$, so the partial derivative in equation~(\ref{eomx}) can be replaced with an ordinary derivative and we deduce that $v_y$ depends only on $x$ (a result reminiscent of the Proudman--Taylor theorem).

Since we consider axisymmetric motions of an ideal fluid, the specific angular momentum should also be preserved under the transformation. The local version of this quantity (divided by the radius of the reference orbit around which the local model is constructed) is the canonical $y$-momentum $2\Omega x+u_y=(2\Omega-S)x+v_y$, which takes into account the angular momentum associated with the rotation of the frame of reference. In order for this quantity to be preserved under the transformation, we require
\begin{equation}
  (2\Omega-S)x+v_y(x)=(2\Omega-S)x_0.
\label{am}
\end{equation}
It follows that $x_0$ is a function of $x$ only, and vice versa, so the Jacobian simplifies to
\begin{equation}
  J=\begin{vmatrix}
    \dfrac{\dd x}{\dd x_0}&0\\[8pt]
    \dfrac{\p z}{\p x_0}&\dfrac{\p z}{\p z_0}
  \end{vmatrix}=\f{\dd x}{\dd x_0}\f{\p z}{\p z_0}.
\end{equation}
A physical interpretation of this result is that the horizontal displacement of the fluid is independent of height. The first factor in this expression for $J$ is the reciprocal of that by which the surface density changes:
\begin{equation}
  \f{\Sigma}{\Sigma_0}=\f{\dd x_0}{\dd x},
\end{equation}
so that the vertically integrated mass element $\Sigma\,\dd x=\Sigma_0\,\dd x_0$ is preserved. The second factor in the expression for $J$ relates to the vertical rearrangement of fluid within each vertical column.

Differentiating equation~(\ref{am}) with respect to $x$ and dividing by $\Sigma$, we find
\begin{equation}
  \f{2\Omega-S+\dfrac{\dd v_y}{\dd x}}{\Sigma}=\f{2\Omega-S}{\Sigma_0}.
\label{pv}
\end{equation}
This is equivalent to the condition that the potential vorticity (also known as vortensity in the literature on astrophysical discs) is preserved, as it would be in a 2D flow. The 3D solutions that we consider are special, being barotropic and having a purely vertical vorticity that is independent of $z$. The equivalence of angular-momentum conservation and vorticity preservation occurs because we consider axisymmetric flows.

Using equation~(\ref{am}) to substitute for $v_y$ in equation~(\ref{eomx}), we find
\begin{equation}
  \kappa^2(x-x_0)=-\f{\dd\psi}{\dd x}.
\end{equation}
Differentiating with respect to $x$ gives
\begin{equation}
  \kappa^2\left(1-\f{\Sigma}{\Sigma_0}\right)=-\f{\dd^2\psi}{\dd x^2},
\end{equation}
which will serve as our equation for both mass and potential vorticity conservation in the analysis to follow.

\section{Isothermal case}\label{s:iso}

In reality, the thermal behaviour of gases in discs undergoing gravitational instability is not described by adiabatic thermodynamics. In particular, the outer parts of protoplanetary discs are heated by radiation from the central star or other nearby stars. This effect is often modelled as a thermal relaxation process, in which the temperature of the gas relaxes on a timescale $\tau$ towards a target temperature $T_0$ set by the external radiation field.  Both $\tau$ and $T_0$ may depend on radial location within the disc.

If the disc undergoes time-dependent perturbations such as a travelling wave or a growing instability with a timescale comparable to $\tau$, then its thermal behaviour will be intermediate between the adiabatic behaviour described by an ideal fluid model with adiabatic exponent $\gamma$ and an isothermal behaviour in which the temperature is fixed at $T_0$. If, however, we are interested in axisymmetric equilibrium structures as part of a bifurcation sequence, then the time-dependence is strictly absent and the equilibria should really be considered to be isothermal as the outcome of  relaxation to the external radiation field.  The relevant sound speed is then the fixed isothermal sound speed $c_\text{s}=\sqrt{p/\rho}\propto T_0$.

We therefore consider the isothermal system separately. A polytropic (i.e.\ adiabatic) extension to this model is discussed in the appendix. The equations governing the marginally stable system (with $\upartial_t \to 0$) may be summarised as
\begin{subequations}\label{isoeqs1}
\begin{align}
  &\upartial_z\left(w + \phi + \frac{1}{2} \nu^2 z^2\right) = 0,\\ 
  &\nabla^2 \phi = 4 \pi G \rho,\\
  &w = c_\text{s}^2\ln(\rho/\rho_R),\\
  &\upartial_x^2\left(w + \phi + \frac{1}{2} \nu^2 z^2\right) = \kappa^2 \left(\frac{\Sigma}{\Sigma_0} - 1\right), \label{isobaseeq}
\end{align}
\end{subequations}
where $\rho_R$ is the Roche density, $\rho_R = \frac{\nu^2}{4\pi G}$.

As in 2D, the system undergoes a finite-wavelength instability when the surface density becomes greater than some critical value, which we denote $\Sigma_c$. We aim to describe analytically the weakly nonlinear structures existing just before the onset of this instability (should the bifurcation be subcritical). Taking inspiration from the 2D case, we'll derive the nonlinear equation governing the evolution of their surface density modulation. Whilst we seek steady equilibria, taking $\upartial_t = 0$, it's possible to later allow for a slow time dependence, which permits the weakly nonlinear structures to travel radially in the disc. We perform this generalisation in section \ref{timedep}.

We start by non-dimensionalising the equations. We let $h = c_\text{s}/\nu$, and set
\begin{equation}
\phi \to \frac{\phi}{c_\text{s}^2}, \quad w \to \frac{w}{c_\text{s}^2}, \quad \rho \to \frac{\rho}{\rho_R}, \quad k \to k h, \quad x \to \frac{x}{h}, \quad z \to \frac{z}{h}, \quad \Sigma \to \frac{\Sigma}{\rho_R h}.
\end{equation}
In this way, for a Keplerian disc (with $\nu^2 = \kappa^2$), the Toomre parameter $Q = 4/\Sigma_0$. The equations become, after assuming $\nu^2 = \kappa^2$:
\begin{subequations}\label{isoeqs}
\begin{align}
  &\upartial_z\left(w + \phi + \frac{1}{2} z^2\right) = 0,\\ 
  &\nabla^2 \phi = \rho, \\
  &w = \ln\rho, \\
  &\upartial_x^2\left(w + \phi + \frac{1}{2} z^2\right) = \left(\frac{\Sigma}{\Sigma_0} - 1\right).\label{isonondim}
\end{align}
\end{subequations}

In order to search for the bifurcation, we write the uniform background surface density, $\Sigma_0$, as:
\begin{equation}
\Sigma_0 = \Sigma_c + \varepsilon^2 \Sigma_2,
\end{equation}
taking $\Sigma_2$ to be a fixed parameter of the problem. $\Sigma_2$ represents the surface density deviation of the reference state from the state of marginal linear stability. Mathematically, we'll treat $\Sigma_2$ as a bifurcation parameter: as $\Sigma_2$ is varied, the reference state transitions from linearly stable to unstable, and sub- or supercritical equilibria may arise via a pitchfork bifurcation on one side of marginal stability.

We look for solutions consisting of a finite-wavenumber disturbance modulated by a slowly-varying envelope. In this way the solution comprises only a few wavenumbers centred about the most unstable, critical, wavenumber in Fourier space. We therefore propose (with the benefit of some hindsight) the following weakly nonlinear expansion of each variable in the form $Y(x,z,X \equiv \varepsilon x;\varepsilon)$:
\begin{multline}\label{ansatz}
Y(x,z,X;\varepsilon) = Y_{0,0}(z) + \varepsilon Y_{1,1}(X,z)\cos(kx) \\ + \varepsilon^2\left[Y_{0,2}(X,z) + Y_{1,2}(X,z)\sin(kx)+ Y_{2,2}(X,z)\cos(2kx)\right] \\ + \varepsilon^3\left[Y_{1,3}(X,z)\cos(kx) + Y_{2,3}(X,z)\sin(2kx) + Y_{3,3}(X,z)\cos(3kx)\right] + \mathcal{O}(\varepsilon^4).
\end{multline}
The $i$ and $j$ in the $Y_{i,j}$ notation above signpost the wavenumber and order in $\varepsilon$ of each term. We have only included Fourier modes at each order in $\varepsilon$ in the above ansatz which end up coupling with the first-order disturbance, $Y_{1,1}(X,z)\cos(kx)$. It will further become apparent that each $Y_{i,j}(X,z)$ is in fact separable in $X$ and $z$; that is, we may write $Y_{i,j}(X,z) = f(X)g(z)$.

Substituting the ansatz (\ref{ansatz}) into the system of equations (\ref{isoeqs}) yields balances at each order in $\varepsilon$ and integer multiple of wavenumber $k$. We'll find the critical surface density $\Sigma_c$ and wavenumber $k_c$ from the linear, first order system. Finally, imposing that the solution to the 1,3 system be orthogonal to the solution to the 1,1 system yields a solvability condition which determines the equation governing the modulation of $\Sigma$. 

\subsection*{0th Order Problem}

The 0th order equations may be manipulated to give the 1D Lane--Emden equation (modified by an external potential), which governs the vertical structure of self-gravitating polytropic discs:
\begin{equation}
\upartial_z^2 w_{0,0} + \rho_{0,0} + 1 = 0, \qquad
\rho_{0,0} = \exp(w_{0,0}).
\end{equation}
This may be integrated (imposing $\p_z w_{0,0} = 0$ on $z = 0$) to give
\begin{equation}
\frac{1}{2}(\upartial_z w_{0,0})^2 + \exp({w_{0,0}}) + w_{0,0} = \exp{(W)}+W,
\end{equation}
where $W = w_{0,0}(0)$. We may find $W$ by requiring the surface density at this order to be $\Sigma_c$:
\begin{equation}\label{sigmacalc}
\Sigma_c(W) = \int_{-\infty}^{\infty}\rho_{0,0}\dd z = 2\int_{0}^{\infty}\exp(w_{0,0})\dd z = \int_{-\infty}^{W}\frac{\sqrt{2} \ee^x}{\sqrt{\ee^W+W - \ee^x - x}}\dd x.
\end{equation}
Therefore, imposing $w_{0,0} = W(\Sigma_c)$ (where $\Sigma_c$ is to be determined in the 1st order analysis), we may simply integrate the above equations to obtain the solution for the 0th order system.

\begin{figure}[H]\centering
                \includegraphics[width=100mm,angle=0]{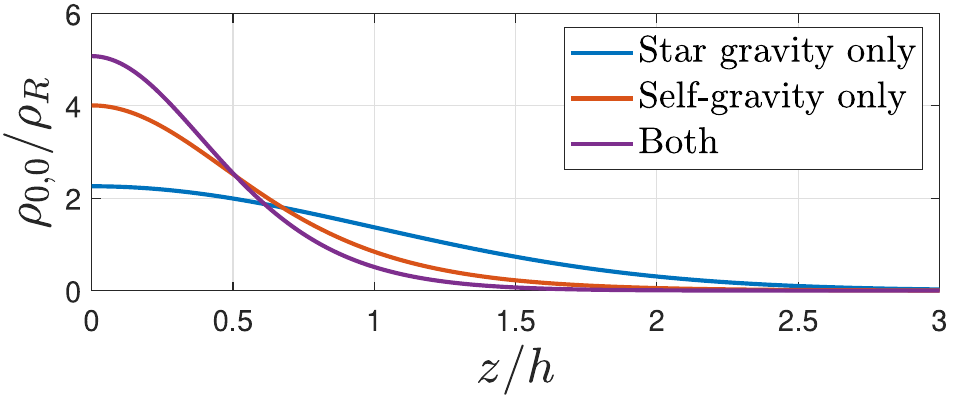}
                \caption{Comparison of the background disc's density distribution when the contributions of the central star's gravity and the disc's local self-gravity are included/neglected. In all cases $\Sigma = \Sigma_c \approx 5.664$ in our non-dimensional units.}
                \label{rho00graph}
\end{figure}

At this order we see the importance of including both the effects of the disc's local self-gravity as well as the global contribution to the gravitational potential from the central star (and possibly the distant parts of the disc). At marginal stability, namely for $\Sigma \approx \Sigma_c$, both effects are comparable in magnitude. Neglecting the star's gravity yields (in our non-dimensional variables)
\begin{equation}
    \rho_{0,0} = \frac{1}{8}\Sigma^2\sech^2\left(\frac{\Sigma z}{4}\right),
\end{equation}
and neglecting self-gravity yields
\begin{equation}
    \rho_{0,0} = \frac{\Sigma}{\sqrt{2\pi}}\exp\left(-\frac{z^2}{2}\right).
\end{equation}
The additional vertical confinement due to both self-gravity and the star's gravity is apparent from the profiles compared in figure \ref{rho00graph}. 

\subsection{1st Order Problem}

The disc will be unstable for $\Sigma > \Sigma_c$, and stable otherwise. Our task at this order is to find $\Sigma_c$ as well as $k_c$, that is, the first wavenumber to become unstable. There will be no solutions at marginal stability (that is, to the system we consider which has $\omega = 0$) for $\Sigma <  \Sigma_c$, two for $\Sigma >  \Sigma_c$, and one (a double root in $k$-space) for $\Sigma = \Sigma_c$. The surface density at which each wavenumber becomes unstable is shown in figure \ref{Scrit} (left). The minimum of the graph corresponds to $\Sigma_c$, the minimum surface density required for linear instability.

The linear system involves perturbations which are proportional to the fractional surface density perturbation $\sigma_{1,1}(X) = \int \rho_{1,1}\dd z/\Sigma_c$ (which is to be determined later as the solution of the nonlinear Klein--Gordon equation (\ref{3DKGE})). In general it's useful to define scaled variables, which we denote with tildes, via
\begin{equation}
    Y_{i,j}(X,z) = \sigma_{1,1}^{j}\tilde{Y}_{i,j}.
\end{equation}
That is, using this definition we scale variables at order $\varepsilon^n$ by $\sigma_{1,1}^n$. 

\subsubsection{The 1,1 system}

\begin{figure}[H]\centering
                \includegraphics[height=72mm,angle=0]{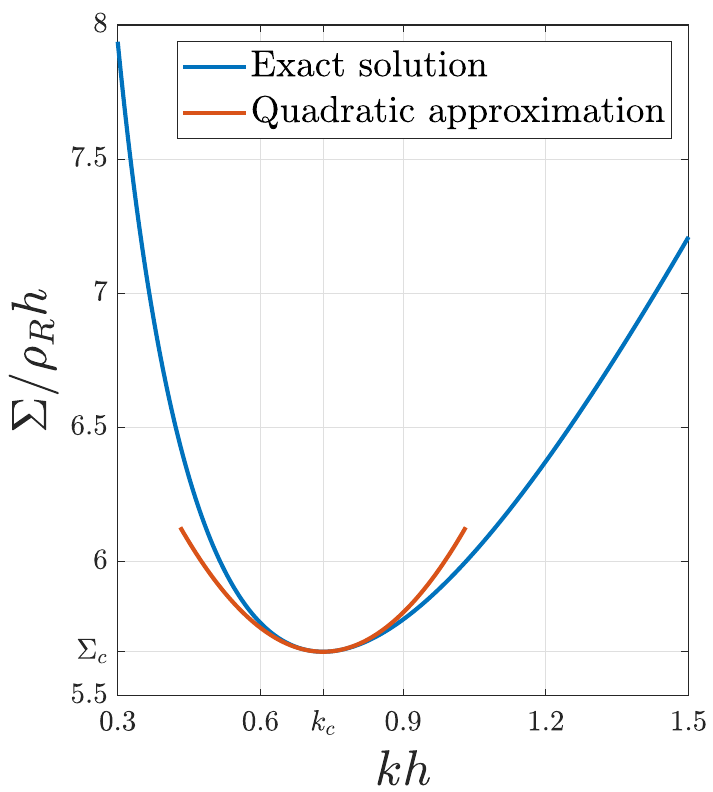}
                \includegraphics[height=72mm,angle=0]{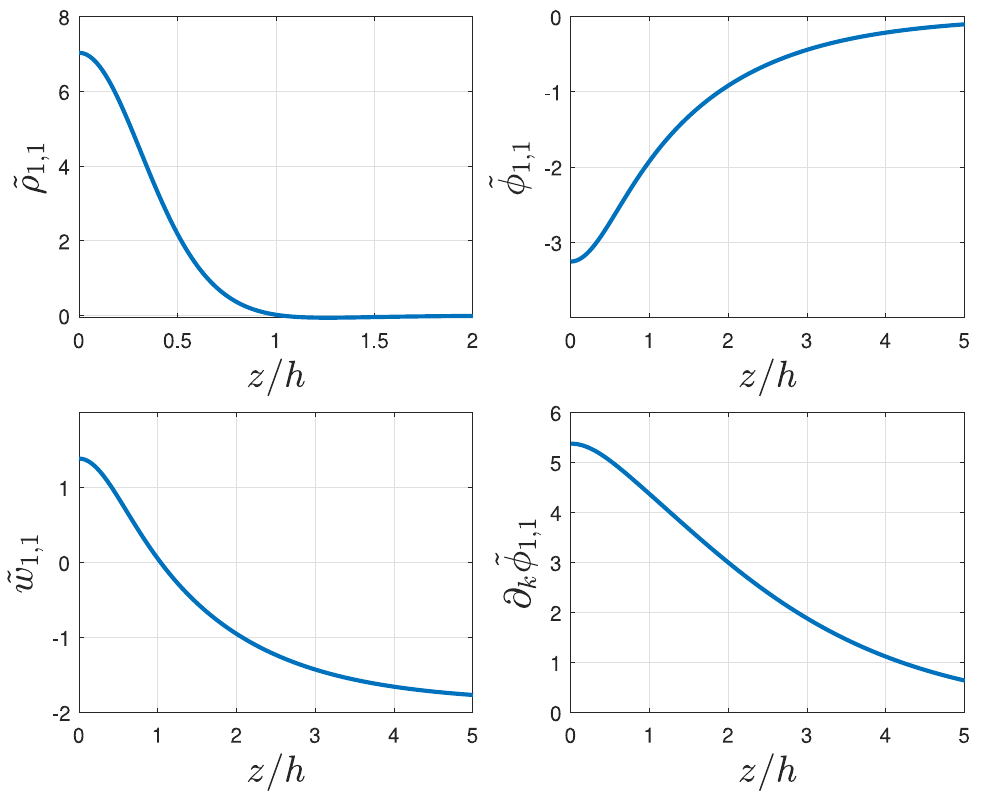}
                \caption{Left: plot of marginal surface densities at which each wavenumber becomes linearly unstable. The smallest surface density for which we reach marginal stability is $\Sigma_c = 5.664\ldots$. The quadratic approximation to this curve near its minimum may be used in an alternative derivation for the spatial variation of the modulating envelope $A(X,T)$, mentioned in section \ref{timedep}. Right: solutions for rescaled variables in the linear system at $\Sigma = \Sigma_c$ and $k = k_c$.}
                \label{Scrit}
\end{figure}

We have at this order
\begin{subequations}\label{11eqs}
\begin{align}
  &\upartial_z^2 \tilde{\phi}_{1,1} - k^2 \tilde{\phi}_{1,1} = \tilde{\rho}_{1,1}, \\
  &\tilde{w}_{1,1} = \tilde{\rho}_{1,1}/\rho_{0,0}, \\
  &-k^2\left(\tilde{w}_{1,1} + \tilde{\phi}_{1,1}\right) = 1,
\label{isonondimscaled1}
\end{align}
\end{subequations}
subject to
\begin{equation}
  \tilde{\phi}_{1,1} \to 0 \text{ as } z \to \infty, \quad \upartial_z \tilde{\phi}_{1,1} = 0 \text{ on } z = 0, \quad \frac{1}{\Sigma_c}\int_{-\infty}^{\infty}\tilde{\rho}_{1,1}\dd z = 1.
\end{equation}
Solving this system for a given wavenumber yields the marginal value of $\Sigma$ at which the system transitions from stable to unstable for that value of $k$ (in considering the steady system we've found the linear mode with $\omega^2 = 0$). We seek the critical value of $k$ corresponding to the minimum of these marginal values, the graph of which is depicted in figure \ref{Scrit} (left).
In order to find $k_c$ and the lowest value of $\Sigma$ necessary for linear instability, we impose the condition $\upartial_k \Sigma = 0$. We differentiate the system (\ref{11eqs}) with respect to $k$, but ignore terms proportional to $\upartial_k\Sigma$. When this system has a solution, we will by necessity have $k = k_c$. The system is
\begin{subequations}
\begin{align}
  &\upartial_z^2 \upartial_k\tilde{\phi}_{1,1} - k^2 \upartial_k\tilde{\phi}_{1,1} = 2 k \tilde{\phi}_{1,1} + \upartial_k\tilde{\rho}_{1,1},\label{dk1,1eom} \\
  &\upartial_k\tilde{w}_{1,1} = \upartial_k\tilde{\rho}_{1,1}/\rho_{0,0}, \\
  &\upartial_k\tilde{w}_{1,1} + \upartial_k\tilde{\phi}_{1,1} = \frac{2}{k^3},
\label{isonondimscaleddk1}
\end{align}
\end{subequations}
with the new boundary conditions
\begin{equation} 
  \upartial_k\tilde{\phi}_{1,1} \to 0 \text{ as } z \to \infty, \quad \upartial_z \upartial_k\tilde{\phi}_{1,1} = 0 \text{ on } z = 0, \quad \frac{1}{\Sigma_c}\int_{-\infty}^{\infty}\rho_{0,0}\upartial_k\tilde{w}_{1,1}\dd z = 0.
\end{equation}
Upon solving, we obtain the solution
\begin{equation}
k_c = 0.73161980708, \quad \Sigma_c = 5.66436370150,
\end{equation}
corresponding to a critical Toomre parameter $Q_c = 0.706169$.

It's noteworthy that the numerical solution for $\partial_k\tilde{\phi}_{1,1}$ wasn't strictly necessary for us to compute. Indeed, we obtain a solvability condition by multiplying (\ref{dk1,1eom}) by $\tilde{\phi}_{1,1}$ and integrating (performing similar algebra to that undertaken in section \ref{scond}). The result may be simplified to
\begin{equation}\label{11scond}
    \int_{-\infty}^{\infty}\tilde{\phi}_{1,1}^2 \dd z = \frac{\Sigma_c}{k_c^4}.
\end{equation}
This may be imposed as an additional boundary condition on the system (\ref{11eqs}) which fixes $k_c$. We solve the remaining nonlinear problem for fixed $k = k_c$.

\subsection{2nd and 3rd Order Problems}

Each of the equations in the system (\ref{isoeqs}) is linear apart from the equation of state $w = \ln \rho$. We may Taylor expand and compare coefficients to find the relations between the densities and enthalpies at each order and wavenumber. The result is
\begin{subequations}
\begin{align}
\frac{\rho_{0,2}}{\rho_{0,0}} & = w_{0,2} + \tfrac{1}{4}w_{1,1}^2,\\
\frac{\rho_{1,2}}{\rho_{0,0}} & = w_{1,2},\\
\frac{\rho_{2,2}}{\rho_{0,0}} & = w_{2,2} + \tfrac{1}{4}w_{1,1}^2,\\
\frac{\rho_{1,3}}{\rho_{0,0}} & = w_{1,3} + w_{1,1}\left(w_{0,2} + \tfrac{1}{2}w_{2,2} + \tfrac{1}{8}w_{1,1}^2\right) \equiv w_{1,3} + F.
\end{align}
\end{subequations}

\subsubsection{The 0,2 system}

The 0,2 system governs the radially homogeneous component of the new equilibrium relative to the marginally stable uniform disc. It depends on the surface density perturbation $\varepsilon^2\Sigma_2$ as well as nonlinear forcing by the 1,1 system.

It's most convenient to solve for the enthalpy $w_{0,2}$ in this system rather than the potential, as the potential can absorb any additive constant and still constitute a solution. The governing equation and boundary conditions are
\begin{align}
&\upartial_z^2\tilde{w}_{0,2} = - \tilde{\rho}_{0,2} = -\rho_{0,0}\left(\tilde{w}_{0,2} + \tfrac{1}{4}\tilde{w}_{1,1}^2\right),\\
&\upartial_z\tilde{w}_{0,2} = 0, \quad \int_{-\infty}^\infty \tilde{\rho}_{0,2}\dd z = \tilde{\Sigma}_2.
\end{align}
In the above equations the parameter $\tilde{\Sigma}_2 \equiv \Sigma_2/\sigma_{1,1}^2$ determines the amplitude of the nonlinear state as a function of the deviation of the surface density from the critical value. Whilst we've yet to determine $\sigma_{1,1}$, and correspondingly $\tilde{\Sigma}_2$, it's worth noting that the solution at this order is linear in $\tilde{\Sigma}_2$, that is, $\tilde{Y}_{0,2} = a(z) + \tilde{\Sigma}_2b(z)$. This form of the solution is what we'll need when we come to derive the equation governing $\sigma_{1,1}$.

\subsubsection{The 1,2 system}

Comparing coefficients of $\varepsilon^2\sin(k_cx)$ yields
\begin{subequations}
\begin{align}
&\upartial_z^2\phi_{1,2} - k_c^2\phi_{1,2} = 2 k_c \phi'_{1,1} + \rho_{0,0}w_{1,2},\\
&- 2k_c (w'_{1,1}+\phi'_{1,1}) - k_c^2(\phi_{1,2} + w_{1,2}) = \sigma_{1,2},
\end{align}
\end{subequations}
which we solve subject to the boundary conditions
\begin{equation}
\upartial_z\phi_{1,2} = 0 \quad \text{on} \quad z = 0, \quad \upartial_z \phi_{1,2} \to 0 \quad \text{as} \quad z \to \infty, \quad \sigma_{1,2} = \frac{1}{\Sigma_c}\int_{-\infty}^{\infty}\rho_{1,2}\dd z.
\end{equation}
The solution may be seen to be directly proportional to $\sigma'_{1,1}(X)$. If we define $\phi_{1,2} = \hat{\phi}_{1,2}\sigma'_{1,1}$, the equations and boundary conditions have the remarkable solution in terms of quantities already known to us:
\begin{equation}
    \hat{\phi}_{1,2} = \frac{1}{k^2}\upartial_k(k^2\tilde{\phi}_{1,1})\Big|_{k_c}, \quad \sigma_{1,2} = \frac{2}{k_c}\sigma'_{1,1}.
\end{equation}
One may also derive a solvability condition from this system, in the same way that we will derive one for the 1,3 system in section \ref{scond}. The result of this analysis yields again equation (\ref{11scond}), the same condition as we found in our analysis of the system involving $\upartial_k \tilde{\phi}_{1,1}$.

\subsubsection{The 2,2 system}

We solve Poisson's equation here for the potential $\tilde{\phi}_{2,2}$, along with PV conservation. Unlike the 0,2 system, this system may be solved directly (for the variables scaled by $\sigma_{1,1}^2$) without explicit knowledge of $\tilde{\Sigma}_2$. The governing equations and boundary conditions are:
\begin{subequations}
\begin{align}
&\upartial_{z}^2\tilde{\phi}_{2,2} - 4k_c^2\tilde{\phi}_{2,2} = \tilde{\rho}_{2,2} = \rho_{0,0}\left(\tilde{w}_{2,2} + \tfrac{1}{4}\tilde{w}_{1,1}^2\right),\\
&- 4k_c^2\left(\tilde{w}_{2,2} + \tilde{\phi}_{2,2}\right) = \tilde{\sigma}_{2,2},\\
&\upartial_z \tilde{\phi}_{2,2} = 0 \quad \text{on} \quad z = 0, \quad \tilde{\phi}_{2,2} \to 0 \quad \text{as} \quad z \to \infty, \quad \tilde{\sigma}_{2,2} = \frac{1}{\Sigma_c}\int_{-\infty}^\infty\tilde{\rho}_{2,2}\dd z.
\end{align}
\end{subequations}

\subsubsection{The 1,3 system}

We need not evaluate the solution at this order numerically; the equations below will allow us to derive the equation governing $\sigma_{1,1}(X)$ via a solvability condition. They read
\begin{subequations}\label{13sys}
\begin{align}
&\upartial_z^2\phi_{1,3} - k_c^2\phi_{1,3} = - 2 k_c \hat{\phi}_{1,2}\sigma''_{1,1} - \tilde{\phi}_{1,1}\sigma''_{1,1} + \rho_{1,3},\label{eom3}\\
&-\frac{1}{k_c^2}\sigma''_{1,1} - k_c^2\left(w_{1,3} + \phi_{1,3}\right) = \sigma_{1,3} -\frac{\Sigma_2}{\Sigma_c}\sigma_{1,1},\label{pv3}\\
&\upartial_z \phi_{1,3} = 0 \quad \text{on} \quad z = 0, \quad \phi_{1,3} \to 0 \quad \text{as} \quad z \to \infty, \quad \sigma_{1,3} = \frac{1}{\Sigma_c}\int_{-\infty}^\infty\rho_{1,3}\dd z.
\end{align}
\end{subequations}

\subsection{Solvability condition}\label{scond}

In order for the 1,3 system to be solvable, that is, for a solution to the system (\ref{13sys}) to exist, we must have that the forcing of this system (which depends on lower order variables) is not resonant with the system's linear differential operator. This can be imposed via a solvability condition, which ensures the forcing is orthogonal to the (resonant) solution of the 1,1 system. If this condition is not met, a slow time evolution must be introduced for the asymptotic series to remain well-ordered (equivalent to the slow time-variability to be introduced in section \ref{timedep}).

The solvability condition fixes the amplitude of the linear solution, $\sigma_{1,1}(X)$, which is the quantity we're interested in. We find it by multiplying (\ref{eom3}) by $\tilde{\phi}_{1,1}$ and integrating:
\begin{equation}
\int_{-\infty}^{\infty}\tilde{\phi}_{1,1}\upartial_{z}^2\phi_{1,3} - k_c^2\tilde{\phi}_{1,1}\phi_{1,3} - \tilde{\phi}_{1,1}\rho_{1,3} \dd z = -\int_{-\infty}^{\infty}\left(2 k_c \hat{\phi}_{1,2} + \tilde{\phi}_{1,1}\right)\tilde{\phi}_{1,1} \dd z \sigma''_{1,1}.
\end{equation}
Integrating by parts and using equations (\ref{11eqs}) and (\ref{pv3}) yields
\begin{align*}
\implies &\int_{-\infty}^{\infty}\phi_{1,3}\upartial_{z}^2\tilde{\phi}_{1,1} - k_c^2\tilde{\phi}_{1,1}\phi_{1,3} - \tilde{\phi}_{1,1}\rho_{1,3} \dd z = - 2 k_c\int_{-\infty}^{\infty}\hat{\phi}_{1,2}\tilde{\phi}_{1,1} \dd z \sigma''_{1,1} - \frac{\Sigma_c}{k_c^4}\sigma''_{1,1},\\
 = &\int_{-\infty}^{\infty}\phi_{1,3}\tilde{\rho}_{1,1} - \tilde{\phi}_{1,1}\rho_{1,3} \dd z\\
= &\int_{-\infty}^{\infty}\rho_{0,0}\tilde{w}_{1,1}\left(\phi_{1,3}+w_{1,3}\right) + \frac{1}{k_c^2}\rho_{1,3} + \rho_{0,0}\tilde{w}_{1,1}F\dd z\\
= & \;\Sigma_c\left(\phi_{1,3}+w_{1,3} + \frac{\sigma_{1,3}}{k_c^2}\right) + \int_{-\infty}^{\infty}\rho_{0,0}\tilde{w}_{1,1}F\dd z\\
= & \;\frac{1}{k_c^2}\Sigma_2\sigma_{1,1} - \frac{1}{k_c^4}\Sigma_c\sigma''_{1,1} + \int_{-\infty}^{\infty} \rho_{0,0}\tilde{w}_{1,1}\tilde{F} \dd z\sigma_{1,1}^3.\\
\end{align*}
Therefore $\sigma_{1,1}$ is given by the solution of the following equation, which is linear in $\Sigma_2$:
\begin{equation}\label{LEint}
    \Sigma_2\sigma_{1,1} + k_c^2\int_{-\infty}^{\infty} \rho_{0,0}\tilde{w}_{1,1}\tilde{F} \dd z\sigma_{1,1}^3 = -2 k_c^3\int_{-\infty}^{\infty} \hat{\phi}_{1,2}\tilde{\phi}_{1,1} \dd z\sigma''_{1,1}
\end{equation}
\begin{equation}\label{LEnum}
\iff \Sigma_2\sigma_{1,1} + 0.8705388149 \Sigma_2\sigma_{1,1} + 0.9689264478\sigma_{1,1}^3 = 9.635534038\sigma''_{1,1}.
\end{equation}
Here we've used the numerical solutions for the systems at first and second order in $\varepsilon$ to evaluate the integrals in equation (\ref{LEint}), exploiting the linear dependence of the 0,2 system on $\tilde{\Sigma}_2$, which gave rise to the second term in equation (\ref{LEnum}) above.

We therefore find (now setting $\varepsilon = 1$ but considering $\Sigma_2$ small) that the surface density perturbation envelope obeys the equation
\begin{equation}\label{GLE}
\sigma''_{1,1} = \alpha_{\text{3D}}^2\sigma_{1,1} - \beta_{\text{3D}}^2\sigma_{1,1}^3,
\end{equation}
for $\alpha_{\text{3D}}^2 = -0.194129231\,\Sigma_2$ and $\beta_{\text{3D}}^2 = 0.100557628$. Equation (\ref{GLE}) admits the stationary equilibrium solution
\begin{equation}
\sigma_{1,1} = \sqrt{2}\frac{\alpha_{\text{3D}}}{\beta_{\text{3D}}}\sech\left(\alpha_{\text{3D}}x\right).
\end{equation}
This corresponds to a weakly nonlinear solution for the fractional surface density, valid for $|\Sigma_2| \ll \Sigma_c$:
\begin{equation}\label{solitonsoln}
\sigma_{1} = \sqrt{2}\frac{\alpha_{\text{3D}}}{\beta_{\text{3D}}}\sech\left(\alpha_{\text{3D}}x\right)\cos{(k_cx)}.
\end{equation}

\section{Time dependence: the linear dispersion relation in 3D discs}\label{timedep}

We may reintroduce time dependence into the equation governing the isothermal modulating envelope by considering the linear dispersion relation in the neighbourhood of the onset of instability. Specifically, we may deduce the coefficient of $\upartial_T^2A$ in this equation from the value of $\frac{\upartial \Sigma}{\upartial \omega^2}$ at $k = k_c$ and $\omega^2=0$. Its value is the ratio between the coefficients of the terms involving $\upartial_T^2A$ and $\Sigma_2 A$ in the resulting nonlinear Klein--Gordon equation.

The time-dependent axisymmetric linear, local equations governing flow perturbations (which are indicated by primed quantities) may be written as
\begin{subequations}
\begin{align}
   &-\ii\omega v'_x - 2 v'_y = - \upartial_x \psi',\\
   &-\ii\omega v'_y + \frac{1}{2} v'_x = 0,\\
   &-\ii\omega v'_z = - \upartial_z \psi',\\
   &-\ii\omega \rho' + \upartial_x\left(\rho_{0,0} v'_x\right) + \upartial_z\left( \rho_{0,0} v'_z\right) = 0,\\
   &\nabla^2 \phi' = \rho',
\end{align}
\end{subequations}
where $\psi' = w' + \phi'$, $w' = \rho'/\rho_0$. 
\begin{align}
    \implies & \upartial_z^2\psi' + (\upartial_z w_0) \upartial_z \psi' + \frac{k^2\omega^2}{1-\omega^2}\psi' + \omega^2(\psi'-\phi') = 0,\\
    &\upartial_z^2\phi' - k^2\phi' = \rho_0(\psi' - \phi').
\end{align}
These must be solved subject to the boundary conditions $\upartial_z \psi' = \upartial_z \phi' = 0$ on $z = 0$, $\phi' \to 0$ as $z \to \infty$, and $\psi'$ polynomially bounded as $z \to \infty$.

\begin{figure}[H]\centering
                \includegraphics[height=70mm,angle=0]{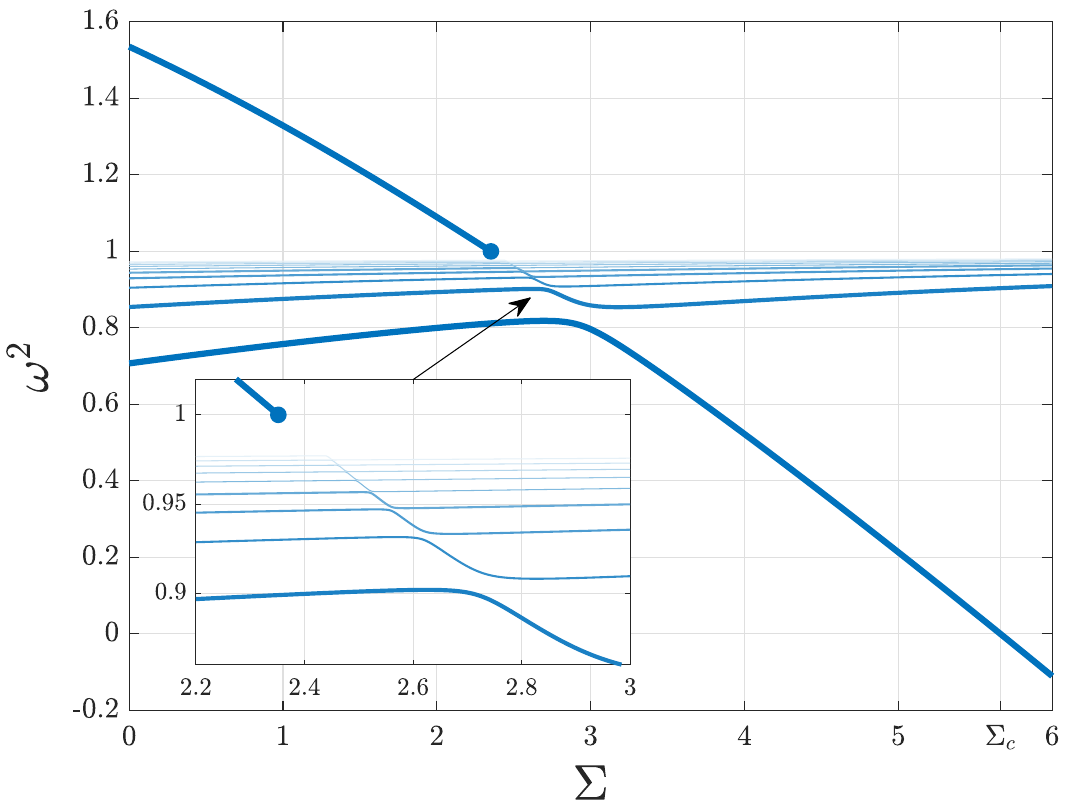}
                \includegraphics[height=70mm,angle=0]{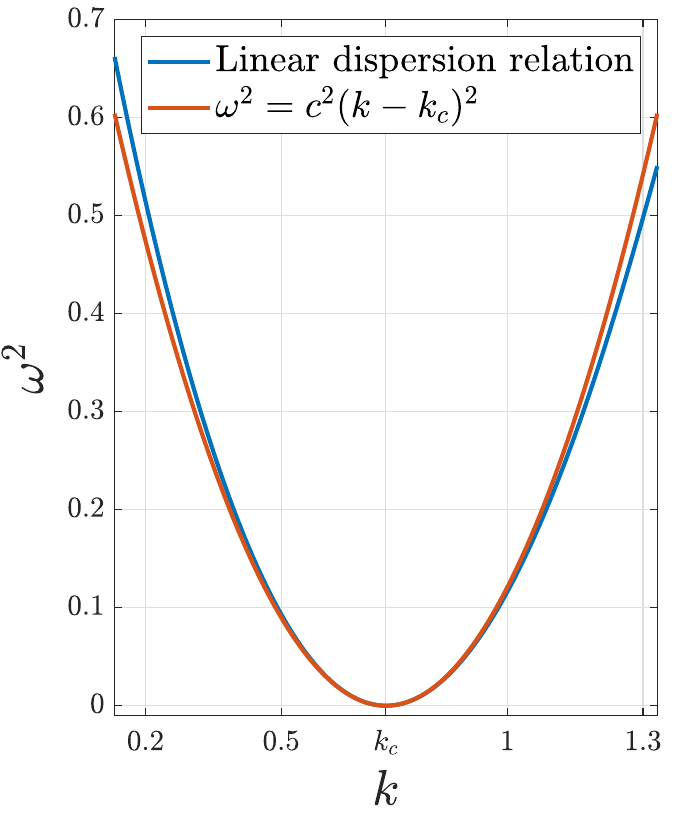}
                \caption{Left: dispersion relation at low frequencies in a 3D isothermal disc at $k = k_c$. There is an infinite number of avoided crossings between the acoustic wave and the inertial waves. Right: Dispersion relation for the mode which becomes unstable for $\Sigma = \Sigma_c$, compared to a quadratic approximation with $c = 1.29549563$.}
                \label{3Ddr}
\end{figure}

This time-dependent 3D system has infinitely many (even) modes which resemble inertial and acoustic waves; however, only one mode ever becomes unstable. Indeed, we only found one solution when we imposed $\omega = 0$ in the analysis in section \ref{s:iso}. The dispersion relation for low frequency modes is depicted in figure \ref{3Ddr} for fixed $k = k_c$ (left), and for fixed $\Sigma = \Sigma_c$ for the mode that becomes unstable to the GI (right). Infinitely many acoustic mode branches exist above the depicted region shown in the left-hand graph. The mode of interest to us crosses the line $\omega^2 = 0$ (as it becomes unstable to the GI), and in the vicinity of the instability the associated disc motion is 2D (with velocity profiles independent of height).

It was not strictly necessary to explicitly include an envelope with long-wavelength modulations in the calculation above, as we could have indirectly deduced the coefficient of $\upartial_X^2 A$ in the governing Klein--Gordon equation from knowledge of the linear system, specifically using the value of $\frac{\upartial^2 \Sigma}{\upartial k^2}$ at $\omega^2 = 0$ and $k = k_c$. This is because the linearised Klein--Gordon equation for $A(X,T)$ describes the dispersion relation in the neighbourhood of the instability onset. We may take a second order (in $\delta k$) expansion of $\Sigma(k)$ about $\Sigma_c$ in the linear dispersion relation (depicted graphically in figure \ref{Scrit} (left)), and upon inverse Fourier transforming, we deduce the coefficient of $\upartial_X^2 A$ relative to the other linear terms in the governing Klein--Gordon equation. Both this approach and the approach taken in section \ref{s:iso} are in exact numerical agreement. 

Similarly, evaluating $\frac{\upartial \Sigma}{\upartial \omega^2} = -3.06928461$ at $\omega^2 = 0$ and $k = k_c$, we find the appropriate time-dependent generalisation of equation (\ref{GLE}) in our non-dimensional system is
\begin{equation}\label{3DKGE}
    \frac{1}{c^2}\frac{\upartial^2 A}{\upartial T^2} -\frac{\upartial^2 A}{\upartial X^2} = -\alpha_{\text{3D}}^2 A + \beta_{\text{3D}}^2 A^3,
\end{equation}
for $c = 1.29549563$, indicating that these solitons may travel with a speed up to around $1.3$ times the sound speed in the disc.

\section{Energy analysis}\label{s:energy}

Our disc model excluded irreversible processes such as heating, and therefore admits a conserved energy. This energy is a functional of permissible flow solutions, and so the task of mapping its topography within the infinite dimensional phase space of solutions is very difficult.

However, restricting to solutions of the form (\ref{sigmaansatz}), we may infer the simplified Lagrangian density of the flow (up to a constant multiple) in terms of the envelope function $A(X,T)$.
\begin{equation}
    \mathcal{L} = \frac{1}{2}\left(\frac{1}{c^2}A_T^2 - A_X^2\right) - V(A),
\end{equation}
for $V(A) = \frac{1}{2}\alpha^2 A^2 - \frac{1}{4}\beta^2 A^4$. The Hamiltonian density is therefore
\begin{equation}
    \mathcal{H} = A_T \frac{\p \mathcal{L}}{\p A_T} - \mathcal{L} = \frac{1}{2}\left(\frac{1}{c^2}A_T^2 + A_X^2\right) + V(A).
\end{equation}
Evaluated for a soliton solution of the form (\ref{3Dsoliton}) which solves equation (\ref{3DKGE}), and travels radially at speed $u$, the above expression simplifies to
\begin{equation}
    \mathcal{H} = \frac{2}{1-u^2/c^2}\frac{\alpha^4}{\beta^2}\sech^2\left(\alpha X'\right)\tanh^2\left(\alpha X'\right),
\end{equation}
where $X' = \frac{\left(X - uT\right)}{\sqrt{1 - u^2/c^2}}$. The solitons therefore have energy density elevated from that of the uniform disc by a factor
\begin{equation}
    \mathcal{E} \propto \frac{\varepsilon^4 \Sigma_2^2}{1-u^2/c^2}\sech^2\left(\alpha X'\right)\tanh^2\left(\alpha X'\right),
\end{equation}
and total energy
\begin{equation}
    E_{\text{tot}} \propto \int \mathcal{E} \dd x = \frac{\sqrt{1-u^2/c^2}}{\varepsilon}\int_{-\infty}^\infty \mathcal{E}\dd X' \propto \frac{\varepsilon^3|\Sigma_2|^{3/2}}{\sqrt{1-u^2/c^2}},
\end{equation}
reminiscent of the total energy of a relativistic particle, $E = \gamma m c^2$. That is, a small but finite minimum energy injection $\Delta E \sim \varepsilon^3|\Sigma_2|^{3/2}$ is necessary to excite one of the soliton solutions, and larger energy perturbations allow for faster travelling solitons.

The soliton solutions may be thought to represent saddle points of the energy functional in phase space, presumably separating the lower energy laminar state and turbulent states, and signposting a nonlinear pathway between the two which requires a finite amplitude perturbation to be traversed. Note that the effective potential $V(A)$ becomes decreasing for large $A$. This is in strong analogy with the role of edge states in the transition to turbulence in pipe flow and other shear flows \citep{2023AnRFM..55..575A}.

\newpage

\section{Results}\label{s:res}

\begin{figure}[H]\centering
                \includegraphics[width=0.495\textwidth,angle=0]{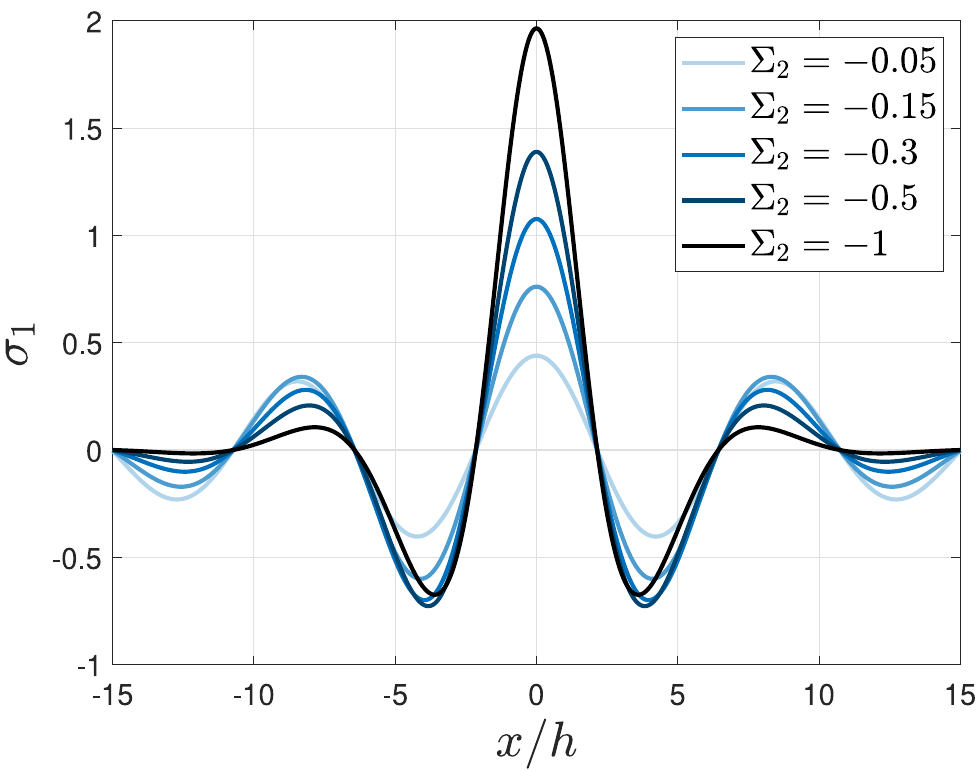}
                \includegraphics[width=0.495\textwidth,angle=0]{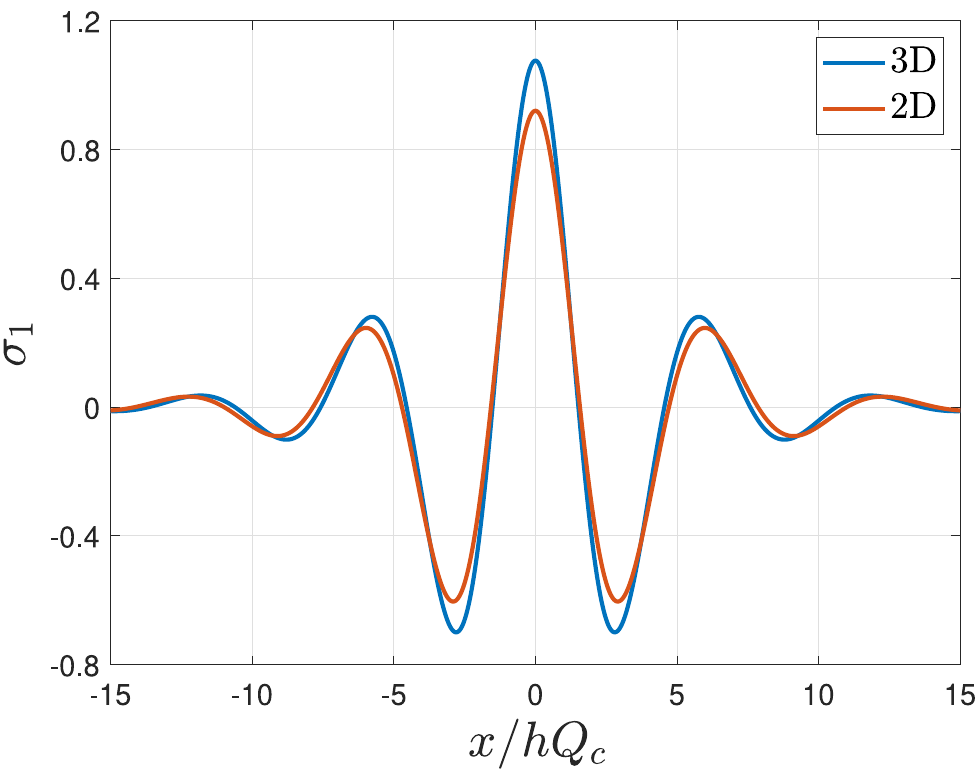}
                \caption{Left: fractional surface density perturbations of 3D stationary isothermal self-gravitating solitons of the form (\ref{solitonsoln}) for various values of $\Sigma_2 = \Sigma_0 - \Sigma_c$. As the background disc's surface density is decreased, stabilising the disc to the GI, the solitons exist with larger amplitude and energy. Right: comparison of 2D and 3D isothermal solitons for $\Sigma_2 = -0.3$, corresponding to $Q/Q_c = 1.056$. Here $Q_c$ denotes the critical value taken by the Toomre parameter at the onset of GI. $Q_c = 1$ in 2D discs, and as discussed in section \ref{s:iso}, $Q_c = 4/\Sigma_c = 0.706$ in a 3D Keplerian isothermal disc. The most unstable wavenumber scales naturally with $Q_c$, motivating the choice of $x$--axis scale.}
                \label{solitoncomp}
\end{figure}

The key result of this work is that for 3D self-gravitating discs with reasonable thermodynamic prescriptions, there is a subcritical bifurcation at the onset of the gravitational instability which gives rise to weakly nonlinear axisymmetric solitary solutions. For isothermal discs, the surface density perturbation of these solitons takes the form
\begin{equation}\label{sigmaansatzR}
    \sigma = \varepsilon A(X,T)\cos(k_c x),
\end{equation}
where $A(X,T)$ is a slowly-varying modulating envelope which depends on the `slow' variables $X = \varepsilon x$ and $T = \varepsilon t$, and obeys the nonlinear Klein--Gordon equation
\begin{equation}\label{3DKGE2}
    \frac{1}{c^2}\frac{\upartial^2 A}{\upartial T^2} -\frac{\upartial^2 A}{\upartial X^2} = -\alpha_{3D}^2 A + \beta_{3D}^2 A^3,
\end{equation}
where $c = 1.29549563$, $\alpha_{\text{3D}}^2 = -0.194129231\Sigma_2$ and $\beta_{\text{3D}}^2 = 0.100557628$. Equation (\ref{3DKGE2}) admits exact soliton solutions of the form
\begin{equation}\label{3Dsoliton}
    A(X-uT) = \sqrt{2}\frac{\alpha_{3D}}{\beta_{3D}}\sech\left(\frac{\alpha_{3D}\left(X - uT\right)}{\sqrt{1 - u^2/c^2}}\right),
\end{equation}
for arbitrary $|u|<c$. Reintroducing dimensions, we see that isothermal solitons may travel radially in the disc at speeds less than around $1.3c_\text{s}$. Figure \ref{solitoncomp} (left) depicts the fractional surface density perturbations for isothermal stationary solitons of the form (\ref{3Dsoliton}) for various values of $\Sigma_2 = \Sigma_0 - \Sigma_c$ (having set $\varepsilon = 1$). Whilst the larger values of $\Sigma_2$ included extend the solution beyond the weakly nonlinear regime, we nevertheless see reasonable qualitative agreement with the large amplitude nonlinear 2D solitons depicted in \citet[fig. 1]{2022ApJ...934L..19D}. Figure \ref{solitoncomp} (right) compares the 3D soliton with the 2D solution from equation (\ref{2Dsoliton}) in the isothermal case. Having scaled the $x-$coordinate by $h Q_c$, the solutions appear remarkably similar.

\begin{figure}[H]\centering
               \includegraphics[width=\textwidth,angle=0]{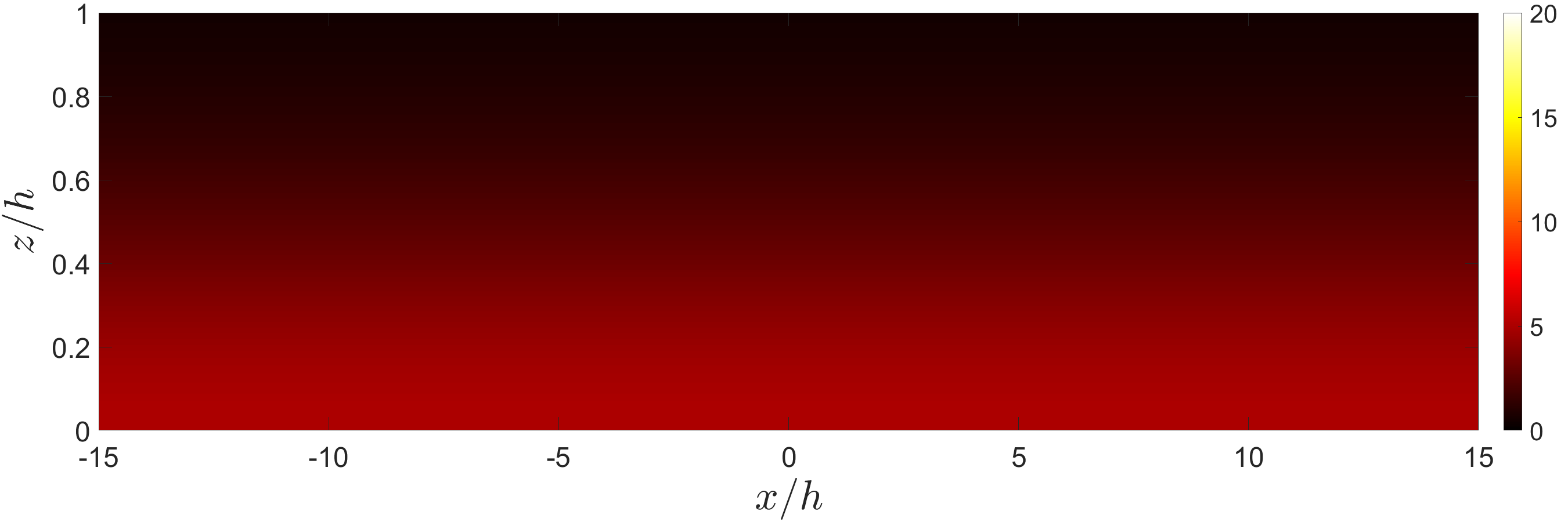}
               \includegraphics[width=\textwidth,angle=0]{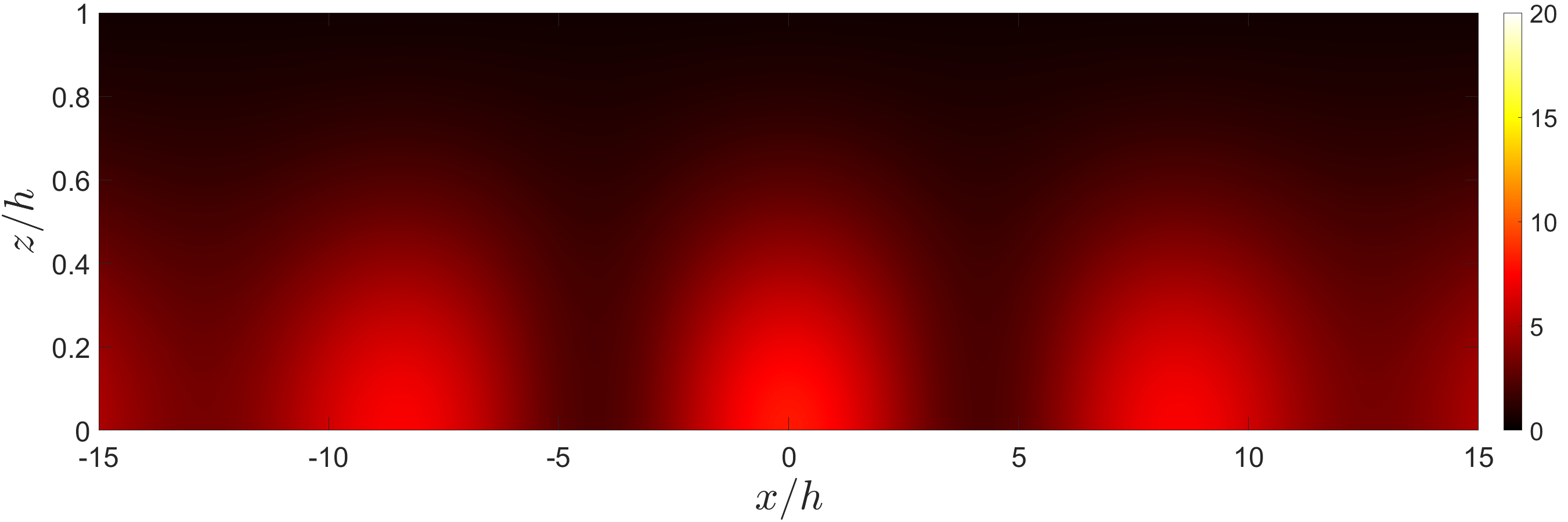}
               \includegraphics[width=\textwidth,angle=0]{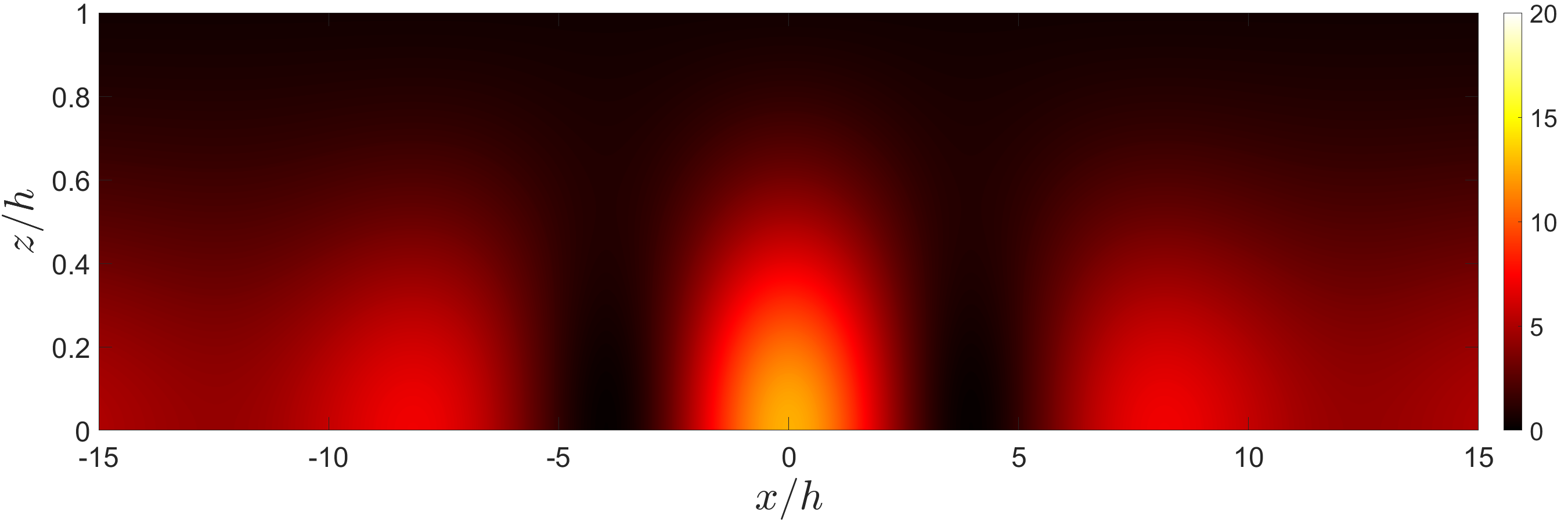}
                \includegraphics[width=\textwidth,angle=0]{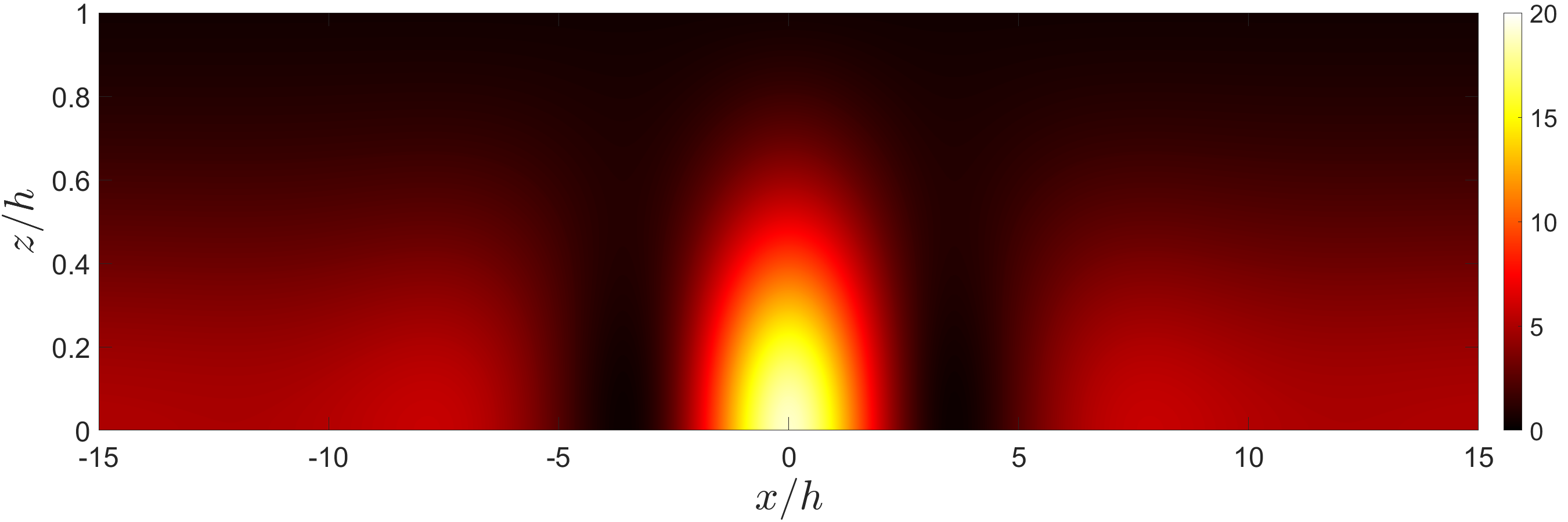}
                \caption{Steady subcritical 3D isothermal soliton density structures for $\Sigma_2 = 0$ (corresponding to a uniform disc at marginal linear stability), $\Sigma_2 = -0.05$, $-0.3$ and $-1$ respectively.}
                \label{fig3dstot2}
\end{figure}

Figure \ref{fig3dstot2} shows the cross-section of the total density structure within the disc for isothermal stationary soliton solutions of increasing amplitude. At larger amplitudes, the solitons resemble axisymmetric rings, with evacuated regions of lower density either side.

We showed further in section \ref{s:energy} that these solitons have energy elevated from that of the uniform disc by a factor
\begin{equation}
    E_{\text{tot}} \propto \frac{\varepsilon^3|\Sigma_2|^{3/2}}{\sqrt{1-u^2/c^2}},
\end{equation}
indicating that a small but finite energy injection $\Delta E \sim \varepsilon^3|\Sigma_2|^{3/2}$ is necessary to reach the soliton solutions from the uniform state.

\begin{figure}[H]\centering
                \includegraphics[width=0.495\textwidth,angle=0]{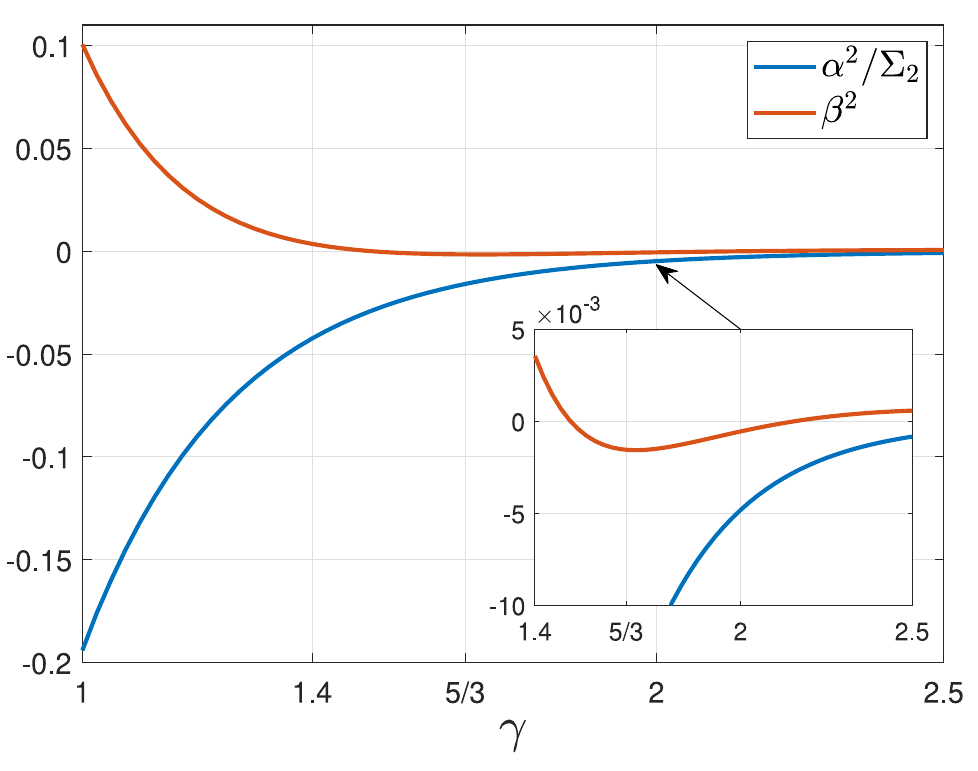}
                \includegraphics[width=0.495\textwidth,angle=0]{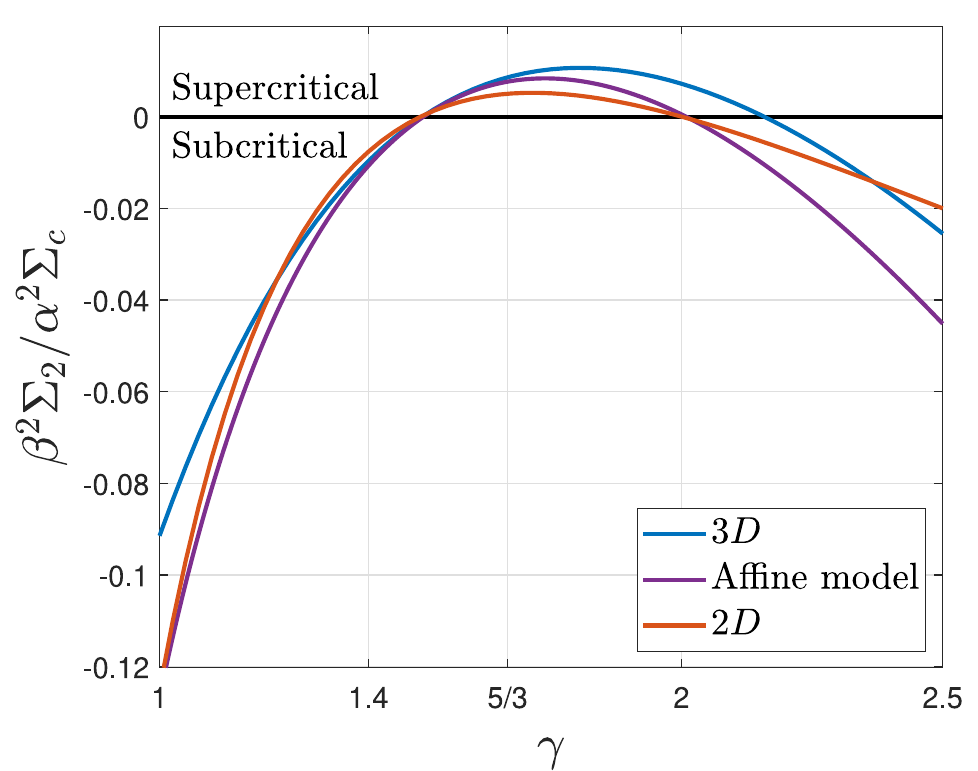}
                \caption{Left: 3D coefficients of the nonlinear Klein--Gordon equation as a function of adiabatic index $\gamma$. Right: criticality of the GI bifurcation in 3D, 2D, and using the affine `2.5D' disc model of \citet{2018MNRAS.477.1744O}.}
                \label{lcoeffs}
\end{figure}

In the appendix, we outline the extension of the isothermal calculation to a polytropic disc. As in the 2D case, the coefficients of the nonlinear Klein--Gordon equation depend on the adiabatic index of the gas which comprises the disc. This dependence is shown in figure \ref{lcoeffs}. Most important is the sign of $\beta^2$ which determines whether the bifurcations are sub- or supercritical, and the ratio $\alpha^2/\beta^2$, which determines the maximum amplitude of the nonlinear state for a fixed value of $\Sigma_2$.
Figure \ref{lcoeffs} (right) compares the bifurcation criticality in 3D with the result from a 2D calculation with equation of state $P = K\Sigma^\Gamma$ (discussed in section \ref{s:2D}) as well a calculation performed using the affine model\footnote{The affine model (an extension of \citet{2018MNRAS.477.1744O} to include self-gravity) treats the disc as a set of columnar fluid elements interacting through pressure and gravity and derives the dynamics of their positions and scale-heights from Hamilton's principle.} of astrophysical discs \citep[submitted to MNRAS]{2024AFFINE}. All three disc models present a qualitatively similar physical picture, in which smaller, more physically relevant choices for $\gamma$ (as well as larger values) yield a subcritical bifurcation, with a supercritical intermediate interval. Reassuringly, in the limit $\gamma \to 1$ in 3D, we recover the same values for each parameter as those computed in the isothermal case. 

Some notable values of the coefficients shown in figure \ref{lcoeffs} as well as other important parameters are listed in table \ref{tab1} below.
\begin{table}[H]
                \small
                \setlength{\tabcolsep}{6pt}
                \centering
                \tbl{Critical values of $\Sigma_c$, $k_c$, disc semi-thickness $Z_c$ and coefficients of the nonlinear Klein--Gordon equation (\ref{3DKGE2}) for various $n \geqslant 1$.\label{tab1}}
                {\begin{tabular}{cccccccc}\toprule
					$n$ & $\gamma$ & $\Sigma_c$ & $k_c$ & $Z_c$ & $\beta^2\Sigma_2/\alpha^2\Sigma_c$ & $\alpha^2/\Sigma_2$ & $\beta^2$\\ 
					\hline
                                $1$ & $2$ & $16.07$ & $0.2627$ & $2.0239$ & $7.178\times10^{-3}$ & $-4.81\times10^{-3}$ & $-5.55\times10^{-4}$ \\ 
                                $1.5$ & $5/3$ & $11.38$ & $0.3699$ & $1.7371$ & $8.571\times10^{-3}$ & $-0.0159$ & $-1.55\times10^{-3}$ \\ 
                                $2.5$ & $1.4$ & $8.623$ & $0.4865$ & $1.7375$ & $-9.691\times10^{-3}$ & $-0.0425$ & $3.55\times10^{-3}$\\ 
                                $5$ & $1.2$ & $6.993$ & $0.5970$ & $2.1525$ & $-0.04035$ & $-0.0900$ & $0.0254$ \\ 
                                $\infty$ & $1$ & $5.664$ & $0.7316$ & $\infty$ & $-0.09144$ & $-0.194$ & $0.101$ \\ \botrule
                \end{tabular}}
\end{table}
In 3D, the bifurcation is subcritical for $\gamma < 1.50350$ and $\gamma>2.14398$, and supercritical otherwise. It follows that the soliton solutions may indeed exist in linearly stable discs for values of $\gamma \lesssim 1.4$ (corresponding to a warm diatomic gas), and in particular the bifurcation is firmly subcritical for values of $\gamma$ close to $1$ (which we've associated with the slow growth and evolution of the solitary equilibria). These solitary structures could therefore play an important role in the transition to turbulence in self-gravitating discs.

\section{Discussion}\label{s:discussion}

\subsection{Comparison with previous work}

Our study was originally intended as an extension to 3D of the work of \citet{2022ApJ...934L..19D}, and we discuss below the relationship between these two investigations.

During the preparation of this paper, we became aware of the remarkable work of A.~M.~Fridman, V.~L.~Polyachenko and collaborators around 1980, summarized in the book of \citet{1984pgs2.book.....F}. In particular, \citet{1979AZh....56..279M} developed a weakly nonlinear theory of axisymmetric density waves in a 2D disc and derived a nonlinear equation admitting solitary waves with a $\sech$ profile. They found that supersonic travelling solitons were possible when $\tfrac{5}{3}<\Gamma<2$ in the linearly unstable case ($Q<1$) and stationary or subsonic travelling solitons when $\Gamma<\tfrac{5}{3}$ (which they argued is equivalent to $\gamma<\tfrac{3}{2}$) in the linearly stable case ($Q>1$). Subsequently, \citet{1980AZh....57..497P} carried out a related calculation for a 3D disc and found that the subcritical (destabilizing) nonlinear behaviour leading to stationary or subsonic solitons occurs for $\gamma<1.404$. They noted that this critical value of $\gamma$, which is decidedly smaller than that of $\tfrac{3}{2}$ suggested by the 2D theory, is essentially the same as the value ($\gamma=\tfrac{7}{5}$) expected for a diatomic molecular gas (if it is warm enough that the rotational degrees of freedom are excited), meaning that such a gas would be of marginal criticality. However, they noted that if the physical conditions caused the gas to behave isothermally, then the subcritical (destabilizing) nonlinear behaviour might indeed occur.

In our adiabatic 3D calculation, we have found that the critical value of $\gamma$ in a Keplerian disc is in fact $1.5035$, which is essentially the same as that suggested by the 2D theory if the equivalence $\Gamma=2-\tfrac{3}{\gamma}$ is employed, and places the (warm) diatomic gas ($\gamma=\tfrac{7}{5}$) in the subcritical regime. There is at least one significant difference between our calculation and that of \citet{1980AZh....57..497P} that could explain this discrepancy: in their treatment of the vertical structure of the disc, they neglected the important contribution of the central object to the vertical gravity that compresses the disc (cf.\ figure~\ref{rho00graph}). (A similar simplifying assumption was adopted by \citet{1965MNRAS.130...97G} in their analysis of linear stability in a 3D disc.) Furthermore, we have emphasized the strong subcriticality found in the isothermal case and its relevance to GI in star- and planet-forming environments.

\subsection{Possible role of the nonlinear equilibria}

Using an asymptotic expansion for weakly nonlinear solutions close to the onset of instability, we have calculated the beginning of a branch of nonlinear equilibria that bifurcates subcritically from the uniform disc. On general grounds, this branch of equilibria is expected to be unstable, even within the subspace of axisymmetric solutions. The equilibria correspond to saddle points in the dynamical phase space, having a slightly higher energy than that of the uniform disc \citep[cf.][]{2022ApJ...934L..19D}. On the far side of the saddle point in the direction away from the uniform disc is a region of phase space in which the energy is again lower than that of the saddle point and may reach values that are lower than that of the uniform disc. The solutions we have computed can be seen as a gateway to this region of phase space in which non-trivial nonlinear dynamics is possible in the linearly stable regime. Indeed, in the 2D problem for certain values of $\Gamma$, the subcritical branch reaches a saddle-node bifurcation and connects with an upper branch of solutions, presumably stable to axisymmetric perturbations, corresponding to local minima of the energy.

The detailed role of the axisymmetric equilibria in gravitational turbulence remains to be elucidated, but we can appeal to the analogous problem of subcritical transition to turbulence in pipe flow \citep{2023AnRFM..55..575A} or other shear flows. Here the computation of exact coherent states such as nonlinear travelling waves and periodic orbits has been found to illuminate the structure of the dynamical phase space and the transition to turbulence.

Some of the many numerical simulations of GI that have been carried out in shearing boxes in recent years have highlighted the role of axisymmetric structures in the dynamics. In particular,
\citet{2017MNRAS.471..317R} identified axisymmetric features in their 3D simulations of GI and discussed their possible role in a subcritical transition to turbulence, drawing an analogy with the magnetorotational dynamo problem. (The axisymmetric structures that they focused on were large-scale epicyclic modes rather than the solitary density waves that we have considered.) Additionally, \citet{2018MNRAS.477.3683V} identified axisymmetric zonal flows in 2D gravitational turbulence and investigated their role in a self-sustaining process. In their study of GI in irradiated 3D boxes, \citet{2019MNRAS.485..266H} showed that axisymmetric density waves often grow first into a nonlinear regime before undergoing secondary, non-axisymmetric instabilities.

\subsection{Caveats}

There are a few main caveats regarding this work which we point out below. 

Most importantly, our thermodynamic assumptions are oversimplified. In practice, the disc's upper layers are typically warmer than its interior as they're irradiated by the central star, whose radiation is not able to penetrate the optically thick disc interior. This contrasts with our model in which temperature was either constant or decreased with height above the midplane. The upper layers re-emit half of the incident radiation into the disc interior. The disc is able to cool via infrared dust emission (despite not necessarily being optically thin to this emission). It's this balance which determines the disc's dominant thermal structure \citep{1997ApJ...490..368C}. Observational studies corroborate this picture, finding a temperature plateau towards a minimum value near the midplane, accompanied by higher temperatures a few scale-heights above the midplane \citep{2003A&A...399..773D,2024ApJ...964..190L}.

Additionally, the thermal relaxation of perturbations to this background state may be quite slow even in the disc's outer regions. We placed particular emphasis on the isothermal case, noting the slow development of equilibrium structures near to marginal stability. It may be possible however that a slow thermal relaxation would interfere with this process. In order to dispel excess heat, the disc's gas must first impart its excess thermal energy to the dust. The dust may then radiate this energy away more efficiently; however, the disc may not be optically thin to this emission, so in practice the heat may need to radiatively diffuse to the surface layers where it may then escape. Furthermore, several authors have pointed out that infrequent gas-dust collision in the disc's upper layers likely act as a bottleneck for the thermal relaxation of the gas near the surface \citep{2017A&A...605A..30M,2021ApJ...912...56B}. In this way, throughout the whole vertical extent of the disc, thermal relaxation timescales could be up to 10 orbits even at 100 au \citep{2021ApJ...912...56B}. This may mean that in practice the development of the solitary equilibria in the bifurcation sequence may not be a completely isothermal process.

Further, we only considered the case of a Keplerian disc, in which the frequency of vertical test-particle oscillations about the disc's midplane, $\nu$, is equal to the epicyclic frequency $\kappa$. This is a good approximation when the disc is in orbit about a spherical object, whose potential dominates the global potential contribution from the disc. In practice however, for a disc massive enough to be near to the gravitational instability threshold, both $\nu^2$ and $\kappa^2$ will receive important contributions from the disc's potential. The ratio $\nu^2/\kappa^2$ will depend on the precise prescription for the disc's morphology, but should remain close to unity. In general, we don't expect the results of our calculation, for example the range of $\gamma$ for which the bifurcation is subcritical, to depend strongly on a small deviation of $\nu^2/\kappa^2$ from $1$. Indeed, in the extreme case $\nu^2/\kappa^2 = 0$, \citet{1980AZh....57..497P} find that the bifurcation is subcritical for $\gamma < 1.404$, which is comparable to our results for the case $\nu^2/\kappa^2 = 1$ in which the bifurcation is subcritical for $\gamma < 1.5035$.

It should also be pointed out that formally the range of validity of our soliton solutions is confined to a small neighbourhood around the bifurcation point. Even for $\Sigma_2 = -0.3$, the isothermal soliton's fractional surface density perturbation exceeds 1 (cf. figure \ref{solitoncomp}) and so the asymptotic ordering in our weakly nonlinear analysis begins to break down. This problem becomes worse as $\gamma$ is increased towards $1.5035$: indeed the soliton's amplitude diverges for fixed $\Sigma_2$ in this limit.

\section{Conclusions}\label{s:conclusion}

Self-gravitating astrophysical discs which are close to the threshold for gravitational instability often exhibit a turbulent state in which the average value of the Toomre stability parameter $Q$ is greater than the critical value for linear instability, naturally raising the question of how the turbulence is sustained in this regime. Following the 2D work of \citet{2022ApJ...934L..19D}, we studied 3D Keplerian polytropic discs with uniform entropy and potential vorticity in a local model, and found that as in the 2D case, there is a pitchfork bifurcation at the onset of the instability, which is subcritical for adiabatic index $\gamma < 1.50350$ and $\gamma>2.14398$, and supercritical otherwise. 

When the bifurcation is subcritical, weakly nonlinear solitary equilibria (which resemble those shown in figure \ref{fig3dstot2}) exist in the stable disc regime, and may travel radially at speeds up to around $1.3$ times the sound speed in an isothermal disc. These `solitons' constitute radially periodic surface density perturbations modulated by a slowly varying envelope function, which obeys a nonlinear Klein--Gordon equation, namely equation (\ref{3DKGE2}). They are accessible via an ideal fluid flow from a uniform, laminar disc state, and have energy only slightly greater than that of the uniform state. In this way, they may be accessed from the uniform, linearly stable state by finite but small perturbations, and probe the energy landscape within the nonlinear phase space of the GI. The solitons themselves are expected to be unstable to non-axisymmetric (as well as axisymmetric) perturbations, providing a possible nonlinear pathway for a stable laminar disc to reach a more energetically favourable turbulent state. 

Looking forwards, simulations which, rather than adopting a $\beta$-cooling prescription that cools the disc towards the state with $Q = 0$ (which guarantees that the GI is triggered), instead permit the disc to relax thermally towards an isothermal state with $Q \gtrsim Q_c$, could provide further verification for the nonlinear dynamics which we predict close to the onset of the gravitational instability, as well as additional insight the nature of the self-sustaining turbulence.

\section*{Acknowledgements}

This research was supported by the Science and Technology Facilities Council (STFC) through grant ST/X001113/1, and by an STFC PhD studentship (grant number 2750631). The authors would like to thank Hongping Deng for helpful discussions.

\section*{Data availability statement}

The data that support the findings of this study are available from the corresponding author, J. B., upon reasonable request.

\section*{Disclosure statement}

The authors report there are no competing interests to declare.

\appendices
\section{Polytropic case}\label{appx}

In this appendix we will relax the isothermal equation of state, and instead solve the polytropic system, in which the disc has a (complicating) finite vertical extent $Z(x)$. In view of brevity, we give a few details on how the calculation may be undertaken. As before, we assume the disc to be locally isentropic. The governing equations are
\begin{subequations}
\begin{align}
  &\f{\p^2\phi}{\p x^2}+\f{\p^2\phi}{\p z^2}=4\pi G\rho,\\
  &w=(n+1)K\rho^{1/n},\\
  &w+\phi+\f{1}{2}\nu^2z^2=\psi(x),\\
  &\f{\dd^2\psi}{\dd x^2}=\kappa^2\left(\f{\Sigma}{\Sigma_0}-1\right),\label{d2psi}\\
  &\Sigma(x)=\int_{-Z(x)}^{Z(x)}\rho(x,z)\,\dd z.
\end{align}
\end{subequations}
In this analysis we'll consider unmodulated solutions which are periodic in $x$, with periodicity length $L_x=2\pi/k_c$, where $k_c$ is the first unstable wavenumber. As discussed in section \ref{timedep}, we are able to deduce the equation governing the modulating envelope from this analysis and the linear dispersion relation. 

We assume that the solution is reflectionally symmetric in $z$. Equation~(\ref{d2psi}) implies that the horizontal mean of $\Sigma$ is equal to $\Sigma_0$, which is consistent with mass conservation. The equations are to be solved in the region $0<x<L_x$, $-Z(x)<z<Z(x)$. In the vacuum regions above and below the disc, we have $\rho=0$ and $\phi$ satisfies Laplace's equation,
\begin{equation}
  \f{\p^2\phi}{\p x^2}+\f{\p^2\phi}{\p z^2}=0.
\end{equation}
At distances much greater than $L_x$ above and below the disc, the local gravitational field should tend to that generated by a disc of uniform surface density $\Sigma_0$. The relevant boundary conditions are therefore that
\begin{equation}
  \left(\f{\p\phi}{\p x},\f{\p\phi}{\p z}\right)\to(0,\pm2\pi G\Sigma_0)\qquad\text{as}\quad z\to\pm\infty.
\end{equation}
Furthermore, $\phi$ and its gradient must be continuous at the surfaces $z=\pm Z(x)$, and $w$ must vanish at these free surfaces.

It's helpful to non-dimensionalise these equations as in the isothermal case. We define the polytropic velocity-scale $c_n$ and length-scale $h$ via

\begin{equation}
c_n^2 = K \left(\frac{\nu^2}{4 \pi G}\right)^{1/n} = K \rho_R^{\frac{1}{n}}, \quad h = \frac{c_n}{\nu},
\end{equation}
and let
\begin{equation}
\phi \to \frac{\phi}{c_n^2}, \quad w \to \frac{w}{c_n^2}, \quad \rho \to \frac{\rho}{\rho_R}, \quad Z \to \frac{Z}{h}, \quad k \to k h, \quad x \to \frac{x}{h}, \quad z \to \frac{z}{h}, \quad \Sigma \to \frac{\Sigma}{\rho_R h}.
\end{equation}
We introduce the stretched vertical coordinate $\zeta$ in order to map the region occupied by the disc onto a rectangular region. This avoids introducing artificial singularities at the disc's surface in the analysis to follow, and allows for separation of variables in a weakly nonlinear analysis. 
\begin{equation}
  \zeta=\f{z}{Z(x)}.
\end{equation}
We change independent variables from $(x,z)$ to $(x,\zeta)$. One periodic cell of the disc then occupies the region $0<x<L_x$, $-1<\zeta<1$. According to the chain rule, the equations are transformed into
\begin{align}
  &\f{\p^2\phi}{\p x^2}+\left(\f{1+Z'^2\zeta^2}{Z^2}\right)\f{\p^2\phi}{\p\zeta^2}+\left(\f{2Z'^2}{Z^2}-\f{Z''}{Z}\right)\zeta\f{\p\phi}{\p\zeta}\nonumber -\f{2Z'}{Z}\zeta\f{\p^2\phi}{\p x\p\zeta}= \rho,\label{poisson_zeta}\\
  &w=(n+1)\rho^{1/n},\\
  &\left(w+\phi+\f{1}{2}Z^2\zeta^2\right)=\psi(x),\\
  &\f{\dd^2\psi}{\dd x^2}=\left(\f{\Sigma}{\Sigma_0}-1\right),\\
  &\Sigma(x)=2Z\int_{0}^1\rho(x,\zeta)\,\dd\zeta,
\end{align}
where $Z'=\dd Z/\dd x$ and $Z''=\dd^2Z/\dd x^2$.

Exploiting the assumed reflectional symmetry, we solve these equations in the region $\zeta>0$ only, applying the symmetry condition $\p\phi/\p\zeta=0$ at $\zeta=0$. In the vacuum region $\zeta>1$, we instead solve Laplace's equation (equivalent to equation (\ref{poisson_zeta}) with $\rho=0$) for $\phi$. We require continuity of $\phi$ and $\p\phi/\p\zeta$ at $\zeta=1$.

We proceed in the same way as in the isothermal case, namely via a perturbative expansion in each variable of the form
\begin{multline}\label{polyansatz}
Y(x,\zeta;\varepsilon) = Y_{0,0}(\zeta) + \varepsilon Y_{1,1}(\zeta)\cos(kx) + \varepsilon^2\left[Y_{0,2}(\zeta) + Y_{2,2}(\zeta)\cos(2kx)\right] \\ + \varepsilon^3\left[Y_{1,3}(\zeta)\cos(kx) + Y_{3,3}(\zeta)\cos(3kx)\right] + \mathcal{O}(\varepsilon^4).
\end{multline}
The disc exterior may be mapped to the upper half plane via the conformal map
\begin{multline}
    f(z) =  z - \ii\bigg[Z_{0,0} + \varepsilon^2 \left(Z_{0,2} + \frac{1}{2}k Z_{1,1}^2\right) + \varepsilon Z_{1,1} \ee^{\ii k \left(z - \ii Z_{0,0}\right)} \\ + \varepsilon^2 \left(Z_{2,2} + \frac{1}{2}k Z_{1,1}^2\right)\ee^{\ii 2 k \left(z - \ii Z_{0,0}\right)}\bigg] + \mathcal{O}\left(\varepsilon^3\right),
\end{multline}
which maps $\{u + \ii H(u); u \in \mathbb{R}\}$ to $\mathbb{R}$ to third order in $\varepsilon$. Since the map preserves the form of Laplace's equation to third order (and approaches a constant downward translation far above the disc surface) the solution for the potential in the disc exterior at each order is easily derived from here. This may be matched onto the interior solution at the boundary by expressing $x$ and $\zeta$ in terms of the conformal coordinates derived from the map $f(z)$. For brevity, we don't include details of the systems which arise at each order.

As in the isothermal case discussed in section \ref{s:iso}, there is a solvability condition at third order which fixes the amplitude of the nonlinear solution. It's easiest to formulate the solvability condition in terms of Eulerian variables which are functions of $x$ and $z$ (instead of our previously adopted semi-Lagrangian variables which are functions of $x$ and $\zeta$). This induces in general (integrable) singularities on the disc surface however, as the discontinuous derivatives of lower order variables at the surface are Taylor expanded to force the higher order systems.

We denote Eulerian variables with a superscript `$\text{E}$', and adopt a similar perturbative expansion to that in equation (\ref{polyansatz}), where now $Y^{\text{E}}_{i,j}$ is a function of $z$, rather than $\zeta$. The Eulerian expansion is related to the semi-Lagrangian expansion by
\begin{subequations}
\begin{equation}
Y^\text{E}_{1,1} = Y_{1,1} - \frac{Z_{1,1}}{Z_{0,0}}\zeta \frac{\dd Y_{0,0}}{\dd \zeta},
\end{equation}
\begin{equation}
Y^\text{E}_{0,2} = Y_{0,2} - \frac{Z_{0,2}}{Z_{0,0}}\zeta\frac{\dd Y_{0,0}}{\dd \zeta} - \frac{1}{4}\frac{Z_{1,1}^2}{Z_{0,0}^2}\zeta^2 \frac{\dd^2 Y_{0,0}}{{\dd \zeta}^2} - \frac{1}{2}\frac{Z_{1,1}}{Z_{0,0}}\zeta \frac{\dd Y^\text{E}_{1,1}}{\dd \zeta},
\end{equation}
\begin{equation}
Y^\text{E}_{2,2} = Y_{2,2} - \frac{Z_{2,2}}{Z_{0,0}}\zeta\frac{\dd Y_{0,0}}{\dd \zeta} - \frac{1}{4}\frac{Z_{1,1}^2}{Z_{0,0}^2}\zeta^2 \frac{\dd^2 Y_{0,0}}{{\dd \zeta}^2} - \frac{1}{2}\frac{Z_{1,1}}{Z_{0,0}}\zeta \frac{\dd Y^\text{E}_{1,1}}{\dd \zeta}.
\end{equation}
\end{subequations}
(In this way, we may also think of $Y^\text{E}_{i,j}$ as a function of $\zeta$.) The solvability condition at third order is
\begin{equation}\label{polysolve}
    \frac{3}{4}k^3Z_{1,1}^2\left(\phi_{1,1}^\text{E}\big|_{\zeta = 1}\right)^2 - \frac{1}{2}\Sigma_2 \sigma_{1,1}^2 + Z_{0,0}\int_0^1 w^\text{E}_{1,3}\rho^\text{E}_{1,1} - w^\text{E}_{1,1}\rho^\text{E}_{1,3}\dd \zeta = 0,
\end{equation}
where
\begin{equation}
 w^\text{E}_{1,3}\rho^\text{E}_{1,1} - w^\text{E}_{1,1}\rho^\text{E}_{1,3} = w^\text{E}_{1,1}\rho^\text{E}_{1,1}\left[\frac{\left(\frac{1}{n} - 1\right)}{2} \left(\frac{2 \rho^\text{E}_{0,2} + \rho^\text{E}_{2,2}}{\rho_{0,0}}\right) + \frac{\left(\frac{1}{n} - 1\right)\left(\frac{1}{n} - 2\right)}{8} \frac{{\rho^\text{E}_{1,1}}^2}{\rho_{0,0}^2}\right].
\end{equation}
The resulting amplitude of the weakly nonlinear solution (which in this appendix is not modulated by a slowly varying envelope) is quantified by the parameter $\tilde{\Sigma}_2 = {\Sigma_2}/{\sigma_{1,1}^2} \equiv \beta^2 \Sigma_2/\alpha^2$, which may be found by solving (\ref{polysolve}). If $\tilde{\Sigma}_2 < 0$, the solution is subcritical, and signposts an unstable equilibrium structure of finite amplitude in a linearly stable disc. The solution for $\tilde{\Sigma}_2$ is shown in figure \ref{lcoeffs} (right), which compares it to the result from a 2D calculation with equation of state $P = K\Sigma^\Gamma$, having identified $\Gamma = 2 - 3/\gamma$.

We may deduce the coefficient of $\upartial_X^2A$ in the nonlinear Klein--Gordon equation for polytropic self-gravitating solitons by considering the linear dispersion relation (as discussed in section \ref{timedep}). Specifically, the coefficient is related to the value of $\frac{\upartial^2 \Sigma}{\upartial k^2}$ at $\omega^2 = 0$ and $k = k_c$. Evaluating this derivative discretely, we find that the modulating envelope for stationary solitons obeys the equation
\begin{equation}
\f{\upartial^2 A}{\upartial X^2} = \alpha^2 A - \beta^2 A^3,
\end{equation}
where $\alpha^2$ and $\beta^2$ are given as functions of $\gamma$ in figure \ref{lcoeffs} (left), and some notable values are listed in table \ref{tab1}.


\end{document}